\documentclass[aps,prb,twocolumn,superscriptaddress,groupedaddress]{revtex4}  
\usepackage{graphicx}  
\usepackage{dcolumn}   
\usepackage{bm}        
\usepackage{amssymb}   
\usepackage{amsmath}
\usepackage{braket}
\usepackage{hyperref}
\hypersetup{
    colorlinks=true,
    linkcolor=blue,
    citecolor=blue,
    filecolor=magenta,      
    urlcolor=cyan,
}
 
\begin{document}



\title{Analytic ground state wavefunctions of mean-field $p_x + i p_y$ superconductors with vortices and boundaries}
\author{Zhiyuan Wang}
\affiliation{Department of Physics and Astronomy, Rice University, Houston, Texas 77005,
  USA}
\email{zhiyuan.wang@rice.edu}
\author{Kaden R.~A. Hazzard}
\affiliation{Department of Physics and Astronomy, Rice University, Houston, Texas 77005,
USA}
\affiliation{Rice Center for Quantum Materials, Rice University, Houston, Texas 77005, USA}
\date{\today}

\begin{abstract}
We study Read and Green's mean-field model of the spinless $p_x+ip_y$ superconductor [N.~Read and D.~Green, Phys. Rev. B {\bf 61}, 10267 (2000)] at a special set of parameters where we find the analytic expressions for the topologically degenerate ground states and the Majorana modes, including in finite systems with edges and in the presence of an arbitrary number of vortices. The wavefunctions of these ground states are similar~(but not always identical) to the Moore-Read Pfaffian states proposed for the $\nu=5/2$ fractional quantum Hall system, which are interpreted as the $p$-wave superconducting states of composite fermions. The similarity in the long-wavelength universal properties is expected from previous work, but at the special point studied herein the wavefunctions are exact even for short-range, non-universal properties. As an application of these results, we show how to obtain the non-Abelian statistics of the vortex Majorana modes by explicitly calculating the analytic continuation of the ground state wavefunctions when vortices are adiabatically exchanged, an approach different from the previous one based on universal arguments. Our results are also useful for constructing particle number-conserving~(and interacting) Hamiltonians with exact projected mean-field states. 
\end{abstract}

\maketitle

\section{Introduction}
In recent years, topological superconductors with $p$-wave pairing symmetry have attracted tremendous attention in condensed matter~\cite{TPSC-RMP,Maj_CM1,Maj_CM2} and cold atom physics~\cite{Maj_coldatom1,Maj_coldatom_QC}. Among the most interesting physical properties is that localized Majorana edge modes appear near system boundaries and in the core of vortices, leading to a topologically protected ground state degeneracy that is robust against small local perturbations. More remarkably, braiding the Majorana modes can give rise to non-Abelian unitary rotations on the space of degenerate ground states, an important example of non-Abelian statistics~\cite{MR1991,Nayak1996,Ivanov} that was first proposed thirty years ago as a theoretical possibility~\cite{Moore1988,Witten1989}. On the practical side, the robustness of the ground state degeneracy and non-Abelian statistics may serve as a noise-tolerant platform for storing and manipulating quantum information, providing a platform for topological quantum computation~\cite{kitaevTC,TPQC-RMP}. Experimental study of topological superconductors has continued to progress~\cite{Exp0,Exp00,Exp1,Exp2,Exp3,Exp4} and a strong evidence of a propagating Majorana zero mode has been reported recently~\cite{ExpZhang}.

A classic example of a one-dimensional~(1D) topological superconductor is Kitaev's Majorana chain~\cite{Kitaev}, a tight-binding model for spinless fermions with nearest neighbor tunneling and $p$-wave pairing. In the topological phase the model exhibits unpaired Majorana edge modes in open boundary conditions, leading to doubly degenerate ground states that can not be distinguished by local measurements. In the topological phase of this model there is a special point $\Delta=t,\mu=0$ at which the Hamiltonian can be represented as a sum of commuting operators and can therefore be easily diagonalized, allowing the quasiparticle energies and wavefunctions of eigenstates to be obtained analytically~\cite{AoP2014}. This makes the model particularly easy to understand, and calculations are greatly simplified at this special point. For example, the non-Abelian statistics of Majorana edge modes was illustrated explicitly at this point in Ref.~[\onlinecite{Alicea2010}]. 

The study of two-dimensional~(2D) topological superconductors was pioneered by Read and Green~\cite{NRead}, who developed the BCS mean-field theory of  a 2D $p_x+ip_y$ superconductor and systematically studied its physical properties. In the topological phase there are localized chiral edge modes~(including a zero-energy Majorana edge mode) near system boundaries and in the core of vortices, giving rise to $2^{M}$ degenerate ground states when there are $2M$ Majorana zero modes. The authors also established the close relationship between a spinless $p_x+ip_y$ superconductor and a $\nu=5/2$ fraction quantum Hall~(FQH) system by identifying the asymptotic~(long-range) behavior of the particle-number-projected BCS wavefunction with the Moore-Read Pfaffian wavefunction~\cite{MR1991} of the FQH state, in translation invariant geometries. The non-Abelian statistics of the vortex Majorana modes was studied one year later~\cite{Ivanov}.

However, unlike Kitaev's Majorana chain, in geometries with boundaries and vortices, the wavefunctions of chiral edge modes were determined only approximately~(typically using WKB approximation) in Read and Green's model, and for the wavefunctions of the degenerate ground states, even an approximate analytic expression is hard to obtain as translation symmetry is broken. For this reason, the study of non-Abelian statistics in Ref.~[\onlinecite{Ivanov}] is based on a heuristic argument of the monodromy of the Majorana zero modes, rather than the explicit calculation of the non-Abelian Berry's matrix from the ground state wavefunctions. 

It is therefore interesting to ask whether it is possible or not to find a special point in Read and Green's model at which all the wavefunctions of degenerate ground states can be exactly obtained in geometries with boundaries and vortices, analogous to the $\Delta=t,\mu=0$ point in Kitaev's Majorana chain. In this paper we show that such a special point does exist in Read and Green's model, once one slightly modifies the boundary Hamiltonian, smoothing out boundaries in a carefully chosen way. At this special point, the task of finding ground state wavefunctions reduces to finding all possible solutions to a first order partial differential equation subject to certain boundary conditions. As a result, the ground state wavefunctions in translation invariant geometries have very simple analytic expressions. Even in geometries with boundaries and vortices, we show that it is possible to slightly modify the Hamiltonian near the boundary so that all the degenerate ground states wavefunctions can be exactly obtained and remain relatively simple. In all geometries we consider the wavefunctions are found to be meromorphic functions of the complex coordinates $z=x+iy$ describing the position of the particles. The ground state wavefunctions with $2M$ vortices have close resemblance to the Moore-Read Pfaffian states with $2M$ quasiholes.

Our result is useful in several ways. First, an outstanding feature of the special point is that all the ground states are annihilated by the local operator $2\partial_{\bar{z}}\psi_z-\psi^\dagger_z$, and this can be used to construct a number-conserving interacting Hamiltonian whose ground states are exactly projected mean-field states. This was done in our previous paper~\cite{ZWang} as a 2D generalization of Ref.~[\onlinecite{SDiehl,NLang}] in which the special point of the Kitaev's Majorana chain is used to construct a 1D number-conserving Hamiltonian with projected mean-field ground states. Secondly, with the analytic wavefunctions at hand, we can use a different approach to study the statistics of vortices, by directly calculating the evolution of the wavefunctions when vortices are exchanged. In this way, we reproduced the result in Ref.~[\onlinecite{Ivanov}] where braiding operators are obtained from universal arguments. Our finding also provide a new example~(and is a rare one in continuum systems) of frustration-free Hamiltonians, namely the ground states of the whole system are also ground states of local Hamiltonians~(even though the local Hamiltonians may not commute with each other). Famous examples of frustration-free models include the Affleck-Kennedy-Lieb-Tasaki~(AKLT) model~\cite{AKLT}, the Rokhsar-Kivelson model~\cite{RKmodel}, and the Kitaev toric code model~\cite{kitaevTC}. The special point is also very suitable for pedagogical purposes, for the essential features~(e.g. topological degeneracy and non-Abelian statistics) of the vortex Majorana modes and chiral edge states can be explicitly calculated by studying the properties of the wavefunctions. 


Our paper is organized as follows. In Sect.~\ref{sec:model} we briefly review the BCS mean-field model of the spinless $p_x+ip_y$ superconductor proposed in Ref.~[\onlinecite{NRead}]. We focus on a special point of this model and motivate our choice to focus on it. In Sect.~\ref{sec:TIG} we introduce the method we will use for the rest of the paper and use it to reproduce previous results for the ground state wavefunctions in translation invariant geometries including an infinite 2D plane and finite size systems with periodic or anti-periodic boundary conditions. In Sect.~\ref{sec:dbwall} we solve the model in a ``double wall'' geometry~(geometry with a finite size in one dimension and hard boundaries) where we obtain the wavefunctions of the ground states and chiral edge excitations. In Sect.~\ref{sec:vortices} we solve the model in geometries with $2M$ vortices. The wavefunctions of the $2^M$ degenerate ground states are obtained. We then use these wavefunctions to study the non-Abelian statistics of the vortex Majorana modes and explicitly demonstrate that the unitary evolution of the degenerate ground states due to vortex exchange is given by the analytic continuation of the ground state wavefunctions. In Sect.~\ref{sec:NC} we show how the special point can be used to construct number-conserving interacting Hamiltonians with exact projected mean-field ground states. We summarize our results in Sect.~\ref{sec:concl}. In Appendix~\ref{appen:norm}  we address a subtlety that appears in calculations at the special point of the mean field superconducting  wavefunction. Specifically, we find that the naive mean-field model yields non-normalizable ground state wavefunctions. We show how the Hamiltonian is regularized in order to obtain finite results.  

\section{The Model}\label{sec:model}
We first recall the BCS mean-field Hamiltonian for spinless fermions with a complex $p$-wave pairing term~\cite{NRead}
\begin{eqnarray}\label{eq:Kmf}
\hat{K}&=&\sum_{\mathbf{k}}[\xi_{\mathbf{k}}\psi^\dagger_{\mathbf{k}}\psi_{\mathbf{k}}+\frac{1}{2}(\Delta^*_{\mathbf{k}}\psi_{-\mathbf{k}}\psi_{\mathbf{k}}+\Delta_{\mathbf{k}}\psi^\dagger_{\mathbf{k}}\psi^\dagger_{-\mathbf{k}})]\nonumber\\
&=&E_{\mathbf{k}} \alpha^\dagger_{\mathbf{k}}\alpha_{\mathbf{k}}
\end{eqnarray}
where $\xi_{\mathbf{k}}=k^2/2m-\mu$, $k=|\mathbf{k}|$,  $E_{\mathbf{k}}=\sqrt{\xi^2_{\mathbf{k}}+|\Delta_{\mathbf{k}}|^2}$, and $\alpha_{\mathbf{k}}$ is the Bogoliubov quasiparticle annihilation operator defined as $\alpha_{\mathbf{k}}=u_{\mathbf{k}}\psi_{\mathbf{k}}-v_{\mathbf{k}}\psi^\dagger_{-\mathbf{k}}$
with $v_{\mathbf{k}}/u_{\mathbf{k}}=-(E_{\mathbf{k}}-\xi_{\mathbf{k}})/\Delta^*_{\mathbf{k}}$ and $|u_{\mathbf{k}}|^2+|v_{\mathbf{k}}|^2=1$. (We set $\hbar=1$.) Ref.~[\onlinecite{NRead}] considered complex $p$-wave pairing, with the function $\Delta_{\mathbf{k}}=i\Delta(k)(k_x-ik_y)$ being an eigenfunction of rotation in $\mathbf{k}$ with angular momentum $L_z=-1$. This model has a topological phase transition at $\mu=0$ with $\mu>0$ being the topological superconducting phase~(the ``weak-pairing'' phase in Ref.~[\onlinecite{NRead}]). 

The function $\Delta(k)$ is generally required to go to zero for large $k$. However, in this paper we find it convenient to set it to a constant $\Delta(k)=\Delta>0$. This will allow us to solve for the mean field wavefunction explicitly, although it also will lead to non-normalizability of the resulting ground state wavefunctions. However, as we show in Appendix \ref{appen:norm}, these states can be regularized by considering them as the limit of a series of normalizable ground states of well-defined mean-field Hamiltonians whose $\Delta(k)\to 0$ for large $k$.

Our other primary simplification of the general $p_x+ip_y$ mean field Hamiltonian is to restrict to the special point $\mu=m\Delta^2/2>0$, where both $E_{\mathbf{k}}$ and $\alpha_{\mathbf{k}}$ are relatively simple:
\begin{eqnarray}
E_k&=&\frac{k^2}{2m}+\frac{m\Delta^2}{2},\nonumber\\
\alpha_{\mathbf{k}}&=&\frac{i(k_x+ik_y)\psi_{\mathbf{k}}-m\Delta\psi^\dagger_{-\mathbf{k}}}{\sqrt{k^2+m^2\Delta^2}}.
\end{eqnarray}
The wavefunction of the ground states with periodic or anti-periodic boundary  condition is already given approximately in Ref.~[\onlinecite{NRead}]. At the special point $\mu=m\Delta^2/2$, one can go further, as in Refs.~[\onlinecite{Dukelsky2010,JSM2010}], to obtain the exact expressions for the ground state
\begin{eqnarray}\label{eq:G_perio}
|G_P\rangle&=&\alpha_{\mathbf{k}=0}\prod_{k_x>0}\alpha_{\mathbf{k}}\alpha_{-\mathbf{k}}|0\rangle\nonumber\\
&\propto&\psi^\dagger_{\mathbf{k}=0}\exp\left(\frac{m\Delta}{2i}\sum_{\mathbf{k}}\frac{\psi^\dagger_{\mathbf{k}}\psi^\dagger_{-\mathbf{k}}}{k_x+ik_y}\right)|0\rangle,\nonumber\\
|G_A\rangle&=&\prod_{k_x>0}\alpha_{\mathbf{k}}\alpha_{-\mathbf{k}}|0\rangle\nonumber\\
&\propto&\exp\left(\frac{m\Delta}{2i}\sum_{\mathbf{k}}\frac{\psi^\dagger_{\mathbf{k}}\psi^\dagger_{-\mathbf{k}}}{k_x+ik_y}\right)|0\rangle,
\end{eqnarray}
where $P,A$ denote periodic and anti-periodic boundary condition, respectively. These exact expressions suggest the remarkable simplification at this point, and as we  will show in the rest of this paper, even in geometries with boundaries and vortices, the wavefunctions of the ground states can still be exactly obtained.

Since our main aim is to exactly solve the model in geometries that break translation invariance, it is helpful to formulate the model in real space. To do so, we rewrite the mean-field Hamiltonian as
\begin{equation}
\hat{K}=\int_{S}\left[ \frac{\nabla \psi_{z}^{\dagger }\cdot \nabla \psi
_{z}}{2m}-(\Delta \psi_{z}\partial_{\bar{z}}\psi_{z}+\mathrm{H.c.})+\mu
\psi_{z}\psi_{z}^{\dagger }\right] d^{2}z,  \label{eq:K_2Dpp}
\end{equation}%
where $S$ denotes a region in the 2D plane with complex
coordinates $z=x+iy$, ${\bar z}$ indicates the complex conjugate, $\partial_{z}=(\partial_{x}-i\partial_{y})/2$, $%
\partial_{\bar{z}}=(\partial_{x}+i\partial_{y})/2$, $d^{2}z=dxdy$, and $%
\psi_{z}$ is the fermionic annihilation operator at position $z$. Here we have removed an irrelevant constant from the chemical potential term. At $\mu=m\Delta^2/2$, we can rewrite $\hat{K}$ as sum of a bulk Hamiltonian and a boundary term $\hat{K}=\hat{K}_{\mathrm{bulk}}+\hat{K}_{\mathrm{bound}}$, with 
\begin{eqnarray}\label{eq:K_bb}
\hat{K}_{\mathrm{bulk}} &=&\int_{S}\frac{1}{2m}(2\partial_{z}\psi_{z}^{\dagger }-m\Delta^*\psi
_{z})(2\partial_{\bar{z}}\psi_{z}-m\Delta\psi_{z}^{\dagger })d^{2}z,  \nonumber\\
\hat{K}_{\mathrm{bound}} &=&-i\oint_{\partial S}\frac{1}{2m}(\psi_{z}^{\dagger
}\partial_{z}\psi_{z}dz+\psi_{z}^{\dagger }\partial_{\bar{z}}\psi_{z}d%
\bar{z}),
\end{eqnarray}%
where $\partial S$ denotes the boundary of $S$. In the following we use natural units 
$2m=m\Delta =1$ for notational simplicity. As we will see in the next section, the factorization of $\hat{K}_{\mathrm{bulk}}$ plays a key role in our analytic solution of this model. 

\section{Ground State Wavefunctions in translation invariant geometries}\label{sec:TIG}
In this section we solve the real space wavefunctions of the ground states in translation invariant geometries. Though this can be easily done by Fourier transforming Eq.~(\ref{eq:G_perio}), here we use a different method which involves solving a differential equation subject to certain boundary conditions. As we will see in the rest of this paper, this new method can be straightforwardly applied to translation non-invariant geometries.
\subsection{An infinite 2D plane}
We first consider an infinite 2D plane in which we set $S=\mathbb{R}^2$ in Eq.~(\ref{eq:K_bb}). In this geometry, the wavefunctions are required to vanish at infinity, so the boundary term $\hat{K}_{\mathrm{bound}}$ in Eq.~(\ref{eq:K_bb}) vanishes. The ground state $|G\rangle$ should therefore minimize $\hat{K}_{\mathrm{bulk}}$. Since $\hat{K}_{\mathrm{bulk}}$ is by construction positive-definite, a sufficient condition for $|G\rangle$ to be a ground state of $\hat{K}$ is if $|G\rangle$ is annihilated by the operator $ 2\partial_{\bar{z}}\psi_{z}-\psi_{z}^{\dagger }$ for $\forall z\in S$. Such a zero-energy ground state with even fermion parity can in general be constructed as~(we will not normalize states throughout this paper)
\begin{equation}
|G\rangle=|G_{[g]}\rangle\equiv\exp \left[ \frac{1}{2}\int_{S}g(z,z^{\prime })\psi
_{z}^{\dagger }\psi_{z^{\prime }}^{\dagger }d^{2}zd^{2}z^{\prime }\right]
|0\rangle ,  \label{eq:Gg}
\end{equation}%
where the two-particle wave function $g(z,z^{\prime })$ satisfies $%
g(z,z^{\prime })=-g(z^{\prime },z)$ and 
\begin{equation}\label{eq:gzz}
2\partial_{\bar{z}}g(z,z') = \delta^{2}(z-z'),  
\end{equation}%
which guarantees that $(2\partial_{\bar{z}}\psi_{z}-\psi_{z}^{\dagger })|G\rangle =0$~\footnote{This can be easily checked using the formula
$$\psi_z|G\rangle=\delta/\delta_{\psi^\dagger_z}|G\rangle,$$
where $\delta/\delta_{\psi^\dagger_z}$ is the functional derivative with respect to the Grassmann field $\psi^\dagger_z$.  }.
Furthermore,  $g(z,z')$ has to satisfy the boundary condition
\begin{equation}\label{eq:bc_TIG1}
g(z,z')\to 0~~\text{for $z\to\infty$, $z'$ fixed,}
\end{equation}
in order for $|G\rangle$ to be a physically reasonable state.

To solve Eq.~(\ref{eq:gzz}) with boundary condition Eq.~(\ref{eq:bc_TIG1}), consider the following identity~(see App.~\ref{app:identities_1})
\begin{equation}\label{eq:deltafunction}
\frac{1}{\pi}\partial_{\bar{z}}\frac{1}{z-z'}=\delta^2(z-z').
\end{equation}
Subtracting Eq.~(\ref{eq:deltafunction}) from Eq.~(\ref{eq:gzz}), we have
\begin{equation}\label{eq:fzz'}
2\partial_{\bar{z}}\left[g(z,z')-\frac{1}{2\pi}\frac{1}{z-z'}\right]=0,
\end{equation}
which means that the function $f(z,z')\equiv g(z,z')-\frac{1}{2\pi(z-z')}$ must be complex analytic in the entire 2D plane. The boundary condition Eq.~(\ref{eq:bc_TIG1}) implies that $f(z,z')$ also has to vanish at infinity, thus from Liouville's theorem, we know that $f(z,z')$ must be identically zero. Therefore,
the ground state of $\hat{K}$ in an infinite 2D plane is
 \begin{eqnarray}\label{eq:G_TIG1}
g(z,z')&=&\frac{1}{2\pi(z-z')},\\
|G\rangle &=&\exp \left[ \frac{1}{4\pi}\int_{S}\frac{1}{z-z^{\prime }}\psi
_{z}^{\dagger }\psi_{z^{\prime }}^{\dagger }d^{2}zd^{2}z^{\prime }\right]
|0\rangle.\nonumber
\end{eqnarray}%
To verify that Eq.~(\ref{eq:G_TIG1}) agrees with Eq.~(\ref{eq:G_perio}) we can use the Fourier transform identity~(see App.~\ref{app:identities_2})
\begin{equation}\label{eq:2DFintegral}
\int \frac{e^{i \mathbf{k}\cdot \mathbf{r}}}{k_x+ik_y}\mathrm{d}^2\mathbf{k}=\frac{2\pi i}{x+i y}.
\end{equation} 

This simple example illustrates the general logic of this paper. As we will see in the following sections, the task of finding the ground state wavefunctions reduces to solving the differential equation (\ref{eq:gzz}) subject to certain boundary conditions. This is the payoff of choosing the special parameter $\mu= m \Delta^2/2$.

\subsection{Periodic and Antiperiodic boundary conditions}
Let us now turn our attention to finite systems with periodic or anti-periodic boundary conditions, with system size $L$ in both $x$ and $y$ directions. We still have to solve the differential equation (\ref{eq:gzz}) but instead of the boundary condition Eq.~(\ref{eq:bc_TIG1}), we want $g(z,z')$ to be (anti-)~periodic in both $z$ and $z'$. For $+-,-+,--$ boundary conditions ($+-$ means periodic in $x$-direction and anti-periodic in $y$-direction, etc.), we can solve Eq.~(\ref{eq:gzz}) using Fourier series
\begin{equation}\label{eq:Fourier}
g(z,z')=\frac{1}{L^2}\sum_{\mathbf{k}}\frac{e^{i\mathbf{k}\cdot(\mathbf{r}-\mathbf{r'})}}{i(k_x+ik_y)},
\end{equation}
where the summation is over all $\mathbf{k}=2\pi (n_x+\eta_x,n_y+\eta_y)/L,~n_{x,y}\in\mathbb{Z}$, $\eta_{x}=0$~(or $1/2$) for periodic~(or anti-periodic) boundary condition in $x$-direction and similarly for $\eta_y$. Evaluating the Fourier series in Eq.~\eqref{eq:Fourier}~(see App.~\ref{app:identities_3}), we get
 \begin{eqnarray}\label{eq:g+-}
g_{+-}(z,z')&=&\frac{1}{2\pi L}\frac{\theta'_1(0|i)}{\theta_1(\frac{z-z'}{L}|i)}\frac{\theta_2(\frac{z-z'}{L}|i)}{\theta_2(0|i)},\nonumber\\
g_{-+}(z,z')&=&\frac{1}{2\pi L}\frac{\theta'_1(0|i)}{\theta_1(\frac{z-z'}{L}|i)}\frac{\theta_4(\frac{z-z'}{L}|i)}{\theta_4(0|i)},\nonumber\\
g_{--}(z,z')&=&\frac{1}{2\pi L}\frac{\theta'_1(0|i)}{\theta_1(\frac{z-z'}{L}|i)}\frac{\theta_3(\frac{z-z'}{L}|i)}{\theta_3(0|i)},
\end{eqnarray}
where $\theta_{i}(z|\tau),~i=1,2,3,4$ are Jacobi theta functions, and $\theta'_{i}(z|\tau)=d\theta_{i}(z|\tau)/dz$. The expressions in Eq.~\eqref{eq:g+-} can be viewed as the (anti-)~periodic versions of Eq.~\eqref{eq:G_TIG1}, since when $z\to 0$, $\theta_1(z|i)\propto z$, while $\theta_{2,3,4}(z|i)$ remain finite, so all the $g(z,z')$ in Eq.~\eqref{eq:G_TIG1} are meromorphic functions of $(z-z')$ having a pole at $z-z'\to 0$ with residue $1/(2\pi)$.

For $++$ boundary condition, the Fourier series Eq.~\eqref{eq:Fourier} becomes ill-defined due to the $\mathbf{k}=0$ mode. In fact, with $++$ boundary condition, the solution to Eq.~\eqref{eq:gzz} does not exist. In this case, instead of considering $|G_{[g]}\rangle$ in Eq.~\eqref{eq:Gg}, we have to consider 
\begin{equation}\label{eq:G++}
|G_{++}\rangle= \psi^\dagger_{\mathbf{k}=0}|G_{[g_{++}]}\rangle,
\end{equation}
i.e. a state with odd fermion parity. Notice that $\{\psi^\dagger_{\mathbf{k}=0},2\partial_{\bar{z}}\psi_z-\psi^\dagger_z\}=0$, therefore, in order that $(2\partial_{\bar{z}}\psi_z-\psi^\dagger_z)|G_{++}\rangle=0$, $g_{++}(z,z')$ should satisfy~\footnote{Eq.~\eqref{eq:g++zz} implies $(2\partial_{\bar{z}}\psi_z-\psi^\dagger_z)|G_{[g_{++}]}\rangle=cL^2\psi^\dagger_{\mathbf{k}=0}|G_{[g_{++}]}\rangle$, leading to $(2\partial_{\bar{z}}\psi_z-\psi^\dagger_z)|G_{++}\rangle= -cL^2(\psi^\dagger_{\mathbf{k}=0})^2|G_{[g_{++}]}\rangle=0$.} 
\begin{equation}\label{eq:g++zz}
2\partial_{\bar{z}}g_{++}(z,z') = \delta^{2}(z-z')+c,  
\end{equation}
where we allow an arbitrary constant $c$ on the right hand side. To solve this equation, we can still use the Fourier series Eq.~\eqref{eq:Fourier} but with the $\mathbf{k}=0$ mode removed
\begin{equation}\label{eq:Fourier++}
g_{++}(z,z')=\frac{1}{L^2}\sum_{\mathbf{k}\neq 0}\frac{e^{i\mathbf{k}\cdot(\mathbf{r}-\mathbf{r'})}}{i(k_x+ik_y)}.
\end{equation}
The constant $c$ turns out to be $-1/L^2$ and the explicit expression for $g_{++}$ is
\begin{equation}
g_{++}(z,z')=\frac{1}{2\pi L}\frac{\theta'_1(\frac{z-z'}{L}|i)}{\theta_1(\frac{z-z'}{L}|i)}+i\frac{y-y'}{L^2}.
\end{equation}
All these results are analogous to the Moore-Read wavefunctions on a torus~\cite{Greiter1991,Greiter1992,Rezayi1996}.

\section{Wavefunctions in a double wall geomerty}\label{sec:dbwall}
\begin{figure}[tbp]
\includegraphics[width=0.6\linewidth]{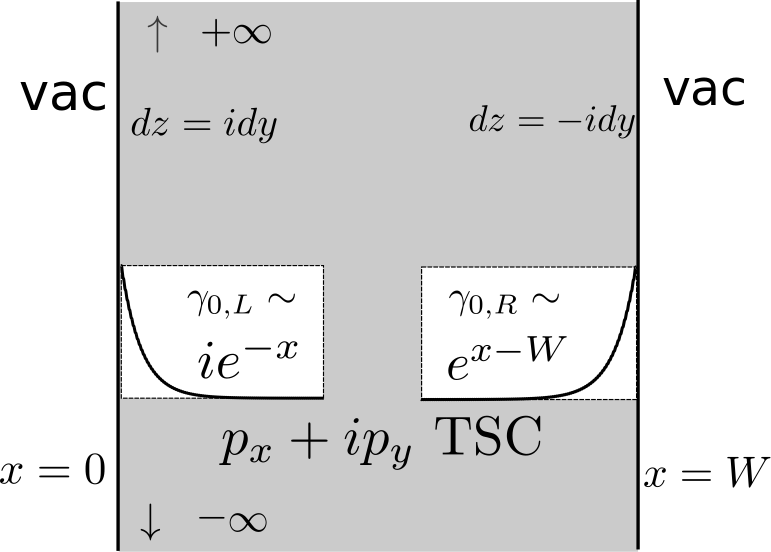}
\caption{Topological superconducting phase of the 2D $p_x+i p_y$ model in double wall geometry. Two straight domain walls are located at $x=0$ and $x=W$, with $x<0$ and $x>W$ being vacuum and $0<x<W$ being the topological superconducting phase. The $y$ dimension is taken to be infinite for simplicity. The insets show the wave functions of Majorana edge modes $\gamma_{0,L},\gamma_{0,R}$ exponentially localized near the domain walls.}
\label{fig:dbwall}
\end{figure}

In this section we calculate the wavefunctions of ground states in a double wall geometry, in which a $p_x+ip_y$ superconductor is separated from the vacuum region by two straight walls located at $x=0$ and $x=W$, as shown in Fig.~\ref{fig:dbwall}. For simplicity, we take the $y$ dimension to be infinite. The Hamiltonian is still given in Eqs.~\eqref{eq:K_2Dpp} and \eqref{eq:K_bb}, with $S$ being the shaded region in Fig.~\ref{fig:dbwall}. Outside the region $S$ we set the chemical potential $\mu=-\infty$ and consequently the wavefunctions vanishes outside $S$ and at the boundary $\partial S$~(near the two walls). As before, the boundary term $\hat{K}_{\mathrm{bound}}$ in Eq.~\eqref{eq:K_bb} vanishes. Naively, we may think that we could solve Eq.~\eqref{eq:gzz} subject to the boundary condition
\begin{equation}\label{eq:bc_dbwall}
g(z,z')= 0~~\text{for $z\in \partial S$ or $z'\in\partial S$}.
\end{equation}
However, we quickly find that such a solution does not exist since a meromorphic function $g(z,z')$ can only have isolated zero points unless identically zero. Indeed, it can be proved that the Hamiltonian Eq.~\eqref{eq:K_2Dpp} does not have a zero energy ground state in this geometry. To get rid of this difficulty, in this paper we slightly modify the Hamiltonian near the boundary so that it allows a zero-energy ground state and our previous method can still be applied. This is similar to the approach in 1D case in Ref.~[\onlinecite{InteractingKitaev}]. We modify the bulk Hamiltonian $\hat{K}_{\mathrm{bulk}}$ in Eq.~\eqref{eq:K_bb} to 
\begin{eqnarray}\label{eq:K'bulk}
\hat{K}'_{\mathrm{bulk}}=\int_{S}\frac{1}{2m}(2\partial_{z}\psi_{z}^{\dagger }-2\partial_z\ln\alpha\ \psi_z-\alpha^2\psi
_{z})\nonumber\\
\times(2\partial_{\bar{z}}\psi_{z}-2\partial_{\bar{z}}\ln\alpha\ \psi_{z}-\alpha^2\psi_{z}^{\dagger })d^{2}z,
\end{eqnarray}
where 
\begin{equation}\label{eq:alpha(x)_dbwall}
\alpha(x)=
\begin{cases}
      &(1-e^{-\lambda x})[1-e^{-\lambda(W-x)}],  0\leq x\leq W, \\
     &0,\ \ \ 0<x \text{ or } W<x,
    \end{cases}
\end{equation}
 $\lambda>0$ is a free parameter, and add a smoothed out boundary term 
\begin{equation}\label{eq:K'_1}
\hat{K}'_1=\int_S [\Delta(z,z')\psi^\dagger_z\psi^\dagger_{z'}+\mu(z,z')\psi^\dagger_z\psi_{z'}+\mathrm{h.c.}]d^2z d^2z',
\end{equation}
where the functions $\Delta(z,z'),\mu(z,z')$ are defined in appendix~\ref{appen:K'}~(they are functions exponentially localized at the boundary and decay exponentially fast as $|z-z'|\to\infty$). Note that as $\lambda\rightarrow \infty$ the Hamiltonian $\hat{K}'_{\mathrm{bulk}}+\hat{K}'_1$, with modifications near the  boundaries, is precisely Eq.~\eqref{eq:K_bb} with hard-wall open boundary condition. The $\hat{K}'_1$ is carefully chosen so that it commutes with $\hat{K}'_{\mathrm{bulk}}$ and therefore they can be chosen to have common eigenstates. In the following we sketch the solution of this model in the main text, with Appendix~\ref{appen:K'} filling in the mathematical details. Using the explicit expressions of $\Delta(z,z'),\mu(z,z')$ in Eq.~\eqref{eq:deltamu_sol}, $\hat{K}'_1$ can be diagonalized as
\begin{eqnarray}\label{eq:K'_1diag}
\hat{K}'_1=\int^{+\infty}_0 k (\gamma^\dagger_{k,L} \gamma_{k,L}+\gamma^\dagger_{-k,R} \gamma_{-k,R}) dk,
\end{eqnarray} 
where the chiral edge modes $\gamma_{k,a}$ are defined in Eq.~\eqref{def:gamma_k}, they satisfy $\gamma^\dagger_{k,a}=\gamma_{-k,a}$, with $a=L,R$, $\gamma_{k,L}$ is localized at the left boundary and $\gamma_{k,R}$ is localized at the right boundary, and they have anticommutation relations
\begin{eqnarray}\label{eq:b_kcommute}
&&\{\gamma_{k,a},2\partial_{\bar{z}}\psi_{z}-2\partial_{\bar{z}}\ln\alpha\ \psi_{z}-\alpha^2\psi_{z}^{\dagger }\}=0,\nonumber\\
&&\{\gamma_{k,a},\gamma_{k',b}\}=\delta(k+k')\delta_{ab}, \text{ for }\forall k,k'\in\mathbb{R},z\in S. \nonumber\\
\end{eqnarray}
In the following we show how to construct zero energy ground states $|G\rangle$ that are annihilated by both $\gamma_{k,L}$ and $\gamma_{-k,R}$ for $k>0$ and $2\partial_{\bar{z}}\psi_{z}-2\partial_{\bar{z}}\ln\alpha\psi_{z}-\alpha^2\psi_{z}^{\dagger }$ for $\forall z$. We use the ansatz state
\begin{eqnarray}\label{eq:Ggalpha}
&&|G\rangle=|G_{[\alpha,g]}\rangle\equiv\nonumber\\
&&\exp \left[ \frac{1}{2}\int_{S}g(z,z^{\prime })\alpha(z)\alpha(z')\psi
_{z}^{\dagger }\psi_{z^{\prime }}^{\dagger }d^{2}zd^{2}z^{\prime }\right]
|0\rangle ,  
\end{eqnarray}
instead of Eq.~\eqref{eq:Gg}. The requirement that $(2\partial_{\bar{z}}\psi_{z}-2\partial_{\bar{z}}\ln\alpha\psi_{z}-\alpha^2\psi_{z}^{\dagger })|G\rangle=0$ still leads to Eq.~\eqref{eq:gzz}, while the condition $\gamma_{k,L}|G\rangle=\gamma_{-k,R}|G\rangle=0$ proves to be equivalent to:
\begin{equation}\label{eq:bc_dbwall-2}
\int^{+\infty}_{-\infty}e^{i k y} g(iy,z')dy=\int^{+\infty}_{-\infty}e^{-i k y} g(W+iy,z')dy=0.
\end{equation}
Appendix~\ref{appen:K'} shows that in this case, the unique solution of $g(z,z')$ satisfying Eqs.~\eqref{eq:gzz} and \eqref{eq:bc_dbwall-2} is $g(z,z')=1/2\pi(z-z')$. There is another ground state with odd fermion parity, which can be obtained by acting the Majorana edge mode $\gamma_{k=0,L}$~(or $\gamma_{k=0,R}$) on $|G_{[\alpha,g]}\rangle$; the resulting state is degenerate because $\gamma_{k=0,L}$ commutes with the Hamiltonian $\hat{K}'_{\mathrm{bulk}}+\hat{K}'_1$. In summary, the two degenerate ground states are
\begin{eqnarray}\label{eq:G_dbwall}
|G_e\rangle&=&\exp \left[ \frac{1}{4\pi}\int_{S}\frac{\alpha(z)\alpha(z')}{z-z^{\prime }}\psi
_{z}^{\dagger }\psi_{z^{\prime }}^{\dagger }d^{2}zd^{2}z^{\prime }\right]
|0\rangle,\nonumber\\
|G_o\rangle&=&\int_S\alpha(z')\psi^\dagger_{z'}d^2 z'|G_e\rangle.
\end{eqnarray} 

It is interesting to point out that the wavefunctions of all those excited states with only chiral edge excitations could also be exactly obtained. For example, if we set 
\begin{equation}\label{eq:g_kk'}
g_{kk'}(z,z')=\frac{1}{2\pi}\frac{1}{z-z'}-\frac{\xi}{2} (e^{-kz-k'z'}-e^{-kz'-k'z})
\end{equation}
in Eq.~\eqref{eq:Ggalpha}, where $\xi\in \mathbb{C}$ and $k,k'\geq 0$, then the state $|G_{kk'}\rangle=|G_{[\alpha,g_{kk'}]}\rangle$ represents a coherent superposition of the ground state and the state with two chiral edge modes with momentum $k$ and $k'$ at the left boundary, i.e.
\begin{equation}\label{eq:G_kk'}
|G_{kk'}\rangle=|G_e\rangle+\xi c_k c_{k'} \gamma^\dagger_{k,L} \gamma^\dagger_{k',L}|G_e\rangle,
\end{equation}
where $c_k,c_{k'}$ are unimportant complex constants. The function $g_{kk'}(z,z')$ also satisfies Eq.~\eqref{eq:gzz} since the additional part $e^{-kz-k'z'}-e^{-kz'-k'z}$ is holomorphic in $z$ and $z'$, so the state $|G_{kk'}\rangle$ minimizes $\hat{K}'_{\mathrm{bulk}}$ as well, i.e. it doesn't excite the bulk Hamiltonian. The condition Eq.~\eqref{eq:bc_dbwall-2} is no longer satisfied for $k$ and $k'$, therefore $|G_{kk'}\rangle$ excites $\hat{K}'_1$, and for $\xi\to \infty$ it is an eigenstate of $\hat{K}'_1$~[as well as the total Hamiltonian, since $\gamma^\dagger_{k,L}$ are eigenmodes of $\hat{K}'_1$ in virtue of Eqs.~\eqref{eq:K'_1diag} and \eqref{eq:b_kcommute}] with energy $E=k+k'$. 


\section{Wavefunctions in an unbounded 2D plane with $2M$ vortices}\label{sec:vortices}
In this section we further generalize our method to vortex geometries and obtain degenerate ground state wavefunctions with arbitrary number of vortices. 
 
\begin{figure}[tbp]
\includegraphics[width=0.6\linewidth]{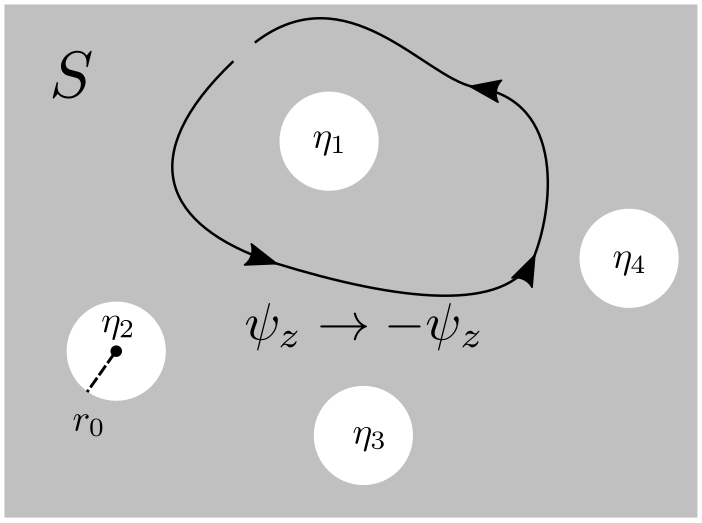}
\caption{\label{fig:vort} Topological superconducting phase in an unbounded 2D plane with $2M$
vortices located at $\protect\eta_1,\protect\eta_2,\ldots, \protect\eta_{2M}$. In this figure we show the $2M=4$ case. In our gauge convention fermion
fields acquire a minus sign on going around each vortex.}
\end{figure}

We consider the geometry shown in Fig.~\ref{fig:vort}, with $2M$ vortices
at $\eta_1,\eta_2,\ldots ,\eta_{2M}$,
respectively. We assume that the distance between any two vortices is large enough, $|\eta_i-\eta_j|\gg 1$, so that chiral edge modes of different vortices do not overlap. Each vortex has a circular core region of radius $r_0\ll 1$ inside which we set the chemical potential to $\mu=-\infty$~(so that wavefunctions vanish in the core region). We use the gauge convention in
which the superconducting order parameter is uniform, and the magnetic field inducing the vortices is implemented by enforcing the  fermion fields to be anti-periodic around each vortex $%
\psi^\dagger_{\theta+2\pi}=-\psi^\dagger_{\theta}$.

The Hamiltonian of the system is formally the same as Eqs.~\eqref{eq:K'bulk} and \eqref{eq:K'_1}, with $S$ denoting the shaded superconducting region in Fig.~\ref{fig:vort}, and the definition of $\alpha$ is replaced by 
\begin{equation}\label{eq:alpha(z)_vort}
\alpha(z)=
\begin{cases}
      &\prod^{2M}_{i=1}\left[1-e^{-\lambda(|z-\eta_i|-r_0)}\right],  z\in S, \\
     &0,\ \ \ z\notin S,
    \end{cases}
\end{equation}
 and again $\Delta(z,z'),\mu(z,z')$ are chosen so that $\hat{K}'_1$ commute with $\hat{K}'_{\mathrm{bulk}}$~(in this case they are functions exponentially localized around vortices, see Appendix~\ref{appen:K'} for their definitions). Using the expressions of $\Delta(z,z'),\mu(z,z')$ in Eq.~\eqref{eq:deltamu_sol_vort}, we can diagonalize $\hat{K}'_1$ as
\begin{eqnarray}\label{eq:K'_1diag_vort}
\hat{K}'_1=\sum^{2M}_{j=1}\sum_{n\in \mathbb{Z}^+} n\gamma^\dagger_{n,j} \gamma_{n,j},
\end{eqnarray} 
where $\gamma_{n,j}$~[see Eq.~\eqref{def:gamma_n} for explicit expression] is the chiral edge mode localized at the $j$-th vortex and they satisfy $\gamma^\dagger_{n,j}=\gamma_{-n,j},1\leq j\leq 2M$, and have anticommutation relations
\begin{eqnarray}\label{eq:b_kcommute_vort}
&&\{\gamma_{n,j},2\partial_{\bar{z}}\psi_{z}-2\partial_{\bar{z}}\ln\alpha\psi_{z}-\alpha^2\psi_{z}^{\dagger }\}=0,\nonumber\\
&&\{\gamma_{n,j},\gamma_{n',j'}\}=\delta_{n,-n'}\delta_{jj'}, \nonumber\\
&&\text{ for }\forall n,n'\in\mathbb{Z},~1\leq j,j'\leq 2M,~z\in S. 
\end{eqnarray}
Therefore, any zero energy ground states $|G\rangle$ should be annihilated by both $\gamma_{n,j}$ for $\forall n>0,1\leq j\leq 2M$ and $2\partial_{\bar{z}}\psi_{z}-2\partial_{\bar{z}}\ln\alpha\psi_{z}-\alpha^2\psi_{z}^{\dagger }$ for $\forall z$. To construct them explicitly, we can still use the ansatz state $|G_{[\alpha,g]}\rangle$ in Eq.~\eqref{eq:Ggalpha}. The requirement that $(2\partial_{\bar{z}}\psi_{z}-2\partial_{\bar{z}}\ln\alpha\psi_{z}-\alpha^2\psi_{z}^{\dagger })|G_{[\alpha,g]}\rangle=0$ still leads to Eq.~\eqref{eq:gzz}, while in this case the condition $\gamma_{n,j}|G\rangle=0$~(for $n>0$) is equivalent to~(see Appendix~\ref{appen:K'} for details):
\begin{equation}\label{eq:bc_vort}
\oint_{|z-\eta_j|=r_0} (z-\eta_j)^{n-1/2}g(z,z')=0.
\end{equation}
In the following subsections we discuss how to find solutions to Eqs.~\eqref{eq:gzz} and \eqref{eq:bc_vort} with increasing number of vortices.

\subsection{Two vortices ($M=1$) case}\label{sec:2vort}
The solution to Eqs.~(\ref{eq:gzz}) and \eqref{eq:bc_vort} with two vortices at $\eta_1,\eta_2$ is determined by the following properties:\\
(a) For the integral in Eq.~\eqref{eq:Gg} to be well-defined, $g(z,z')$ should be multiplied by a phase factor $e^{i\pi}$ when $z$ or $z'$ wind around a vortex~($\eta_1$ or $\eta_2$), in accordance with our gauge convention that the fermion fields are anti-periodic around each vortex.\\
(b) The solution to Eq.~(\ref{eq:gzz}) in an unbounded 2D vortex-free plane is given by $g_0(z,z')=\frac{1}{2\pi}\frac{1}{z-z'}$, see Eq.~\eqref{eq:G_TIG1}. Thus a natural guess for the asymptotic behavior of $g(z,z')$ as $z\rightarrow \infty$ is $g(z,z')\sim 1/z$~(with $z'$ fixed).\\ 
(c) In virtue of Eqs.~(\ref{eq:gzz}) and \eqref{eq:deltafunction}, $g(z,z')$ should have a pole at $z\to z'$ with residue $1/2\pi$.\\
(d) The condition Eq.~\eqref{eq:bc_vort} for $\forall n>0$ indicates that $g(z,z')\sqrt{z-\eta_j}$ is analytic at $z=\eta_j$, as can be seen by considering the Laurent series of $g(z,z')\sqrt{z-\eta_j}$ in the vicinity of $z=\eta_j$.\\
One function satisfying all these conditions is   
\begin{equation}\label{eq:g_12}
g_{1,2}(z,z')=\frac{1}{4\pi}\frac{1}{z-z'}\left[\sqrt{\frac{(z-\eta_2)(z'-\eta_1)}{(z-\eta_1)(z'-\eta_2)}}+(z\leftrightarrow z')\right],
\end{equation}
and it can be checked that this is indeed a solution to Eq.~(\ref{eq:gzz}). To prove that it is the unique solution, we consider the difference between $g_{12}(z,z')$ and $g(z,z')$~(a possibly different solution): 
\begin{equation}\label{eq:dif_g_12}
g(z,z')=g_{1,2}(z,z')+\frac{f(z,z')}{\sqrt{(z-\eta_1)(z-\eta_2)(z'-\eta_1)(z'-\eta_2)}}.
\end{equation}
where $f(z,z')=-f(z',z)$. Now consider $f(z,z')$ as a function of $z$ for a fixed $z'$. Eq.~(\ref{eq:gzz}) requires $f(z,z')$ is a meromorphic function of $z$ and can only have poles at $z=\eta_1,z=\eta_2$ and $z=\infty$. The condition~(c) above requires that $f(z,z')$ should be bounded as $z\to \infty$~(for $z'$ fixed), while the condition~(d) above requires that  $f(z,z')$ should have no pole at $z=\eta_1$ or $z=\eta_2$. In summary, $f(z,z')$ must be analytic everywhere on the complex plane and is bounded when $z\to\infty$, thus from Liouville's theorem, $f(z,z')$ is at most a constant~(possibly dependent on $z'$). Requiring further that $f(z,z')$ is anti-symmetric, we know that $f(z,z')$ must be the zero function. Thus $g_{1,2}(z,z')$ is the unique solution to Eqs.~(\ref{eq:gzz}) and \eqref{eq:bc_vort}, meaning that $|G_e\rangle=|G_{[\alpha,g_{12}]}\rangle$ is the unique ground state with even fermion parity in this geometry. The state with odd fermion parity is obtained by acting either $\gamma_{n=0,j=1}$ or $\gamma_{n=0,j=2}$~(the Majorana zero mode operators localized near $\eta_{1,2}$, respectively) on $|G_e\rangle$, resulting in the state~(we have dropped some constant factors)
\begin{equation}\label{eq:G_o_2vort}
|G_o\rangle=\int_S\sqrt{\frac{\eta_1-\eta_2}{(z-\eta_1)(z-\eta_2)}}\alpha(z)\psi_z^\dagger d^2z|G_e\rangle.
\end{equation}

Similar to the double wall geometry, the wavefunctions of all those excited states with only chiral edge excitations could be exactly obtained. To make the following expressions simpler, let's set $\eta_1=0,\eta_2=\infty$. The ground state wavefunction $g_{1,2}(z,z')$ in Eq.~\eqref{eq:g_12} now simplifies to 
\begin{equation}\label{eq:g_0infty}
g_{0}(z,z')=\frac{1}{4\pi}\frac{1}{z-z'}\left[\sqrt{\frac{z}{z'}}+\sqrt{\frac{z'}{z}}\right].
\end{equation}
Then, analogous to Eq.~\eqref{eq:g_kk'}, we construct the wavefunction
\begin{equation}\label{eq:f_mjnj'}
g_{mn}(z,z')=g_{0}(z,z')-\frac{\xi}{2} [z^{-m-1/2}z'^{-n-1/2}-(z\leftrightarrow z')],
\end{equation}
where $\xi\in \mathbb{C}$ and $m,n\geq 0$, then the resulting state $|G_{mn}\rangle=|G_{[\alpha,g_{mn}]}\rangle$ represents a coherent superposition of the ground state and the state with two chiral edge modes~(localized at $\eta_1$) with momentum $m$ and $n$~(see Appendix~\ref{appen:K'_vort} for derivations), i.e.
\begin{equation}\label{eq:G_mn}
|G_{mn}\rangle=|G_e\rangle+\xi c_m c_{n} \gamma^\dagger_{m,1} \gamma^\dagger_{n,1}|G_e\rangle,
\end{equation}
where $c_m,c_{n}$ are some unimportant complex constants. The functions $g_{mn}(z,z')$ still satisfy Eq.~\eqref{eq:gzz} but $\sqrt{z}g_{mn}(z,z')$ has a pole at $z=0$, which violates Eq.~\eqref{eq:bc_vort} or the condition~(d) above, meaning that $|G_{mn}\rangle$ is not annihilated by $\gamma_{m,1}$ or $\gamma_{n,1}$, and in the limit $\xi\to \infty$ it is an excited state with energy $E=m+n$.

\subsection{Four vortices ($M=2$) case}\label{sec:4vort}
Now we try to obtain the solutions to Eqs.~(\ref{eq:gzz}) and \eqref{eq:bc_vort} in case of 4 vortices located at $\eta_1,\eta_2,\eta_3,\eta_4$, respectively. The conditions (a)-(d) in the previous subsection are still valid in this case, with the only complication being that there are two more vortices now. Directly generalizing Eq.~\eqref{eq:g_12}, we get one solution
\begin{eqnarray}  \label{eq:g_1234}
g_{12,34}(z,z^{\prime })&=&\left[\sqrt{\frac{(z-\eta_1)(z-\eta_2)(z^{\prime
}-\eta_3)(z^{\prime }-\eta_4)}{(z^{\prime }-\eta_1)(z^{\prime
}-\eta_2)(z-\eta_3)(z-\eta_4)}}\right.  \notag \\
&&\left.+(z\leftrightarrow z^{\prime })\vphantom{\frac{1}{2}}\right]\frac{1}{4\pi(z-z^{\prime })}.
\end{eqnarray}
By permuting the indices $1,2,3,4$ we get two different solutions $g_{13,24}(z,z')$ and $g_{14,23}(z,z')$. However, we can prove that~(see Appendix~\ref{appen:2Mdim}) the corresponding states $|G_{12,34}\rangle,|G_{13,24}\rangle,|G_{14,23}\rangle$~[constructed by Eq.~\eqref{eq:Ggalpha}] are linearly dependent
\begin{equation}\label{eq:G_1234}
\eta_{12}\eta_{34}|G_{12,34}\rangle=\eta_{13}\eta_{24}|G_{13,24}\rangle-\eta_{14}\eta_{23}|G_{14,23}\rangle,
\end{equation}
where $\eta_{ij}\equiv \eta_i-\eta_j$, meaning that the degeneracy in the even parity sector is actually two-fold. 
We can construct an orthogonal basis for these states
\begin{eqnarray}\label{eq:basis0011}
|00\rangle&=&\sqrt{N_{00}}(\lambda_{00}|G_{13,24}\rangle+\bar{\lambda}_{00}|G_{14,23}\rangle),\nonumber\\
|11\rangle&=&\sqrt{N_{11}}(\lambda_{11}|G_{13,24}\rangle+\bar{\lambda}_{11}|G_{14,23}\rangle),
\end{eqnarray}
where $N_{00(11)}=\sqrt{\eta_{13}\eta_{24}}\pm\sqrt{\eta_{14}\eta_{23}}$, $\lambda_{00}=\sqrt{\eta_{13}\eta_{24}}/N_{00}$, $\lambda_{11}=\sqrt{\eta_{13}\eta_{24}}/N_{11}$, $\bar{\lambda}_{00}=1-\lambda_{00}$ and $\bar{\lambda}_{11}=1-\lambda_{11}$. These two states form the occupation number basis of the non-local fermions $\chi_1=(\gamma_{0,1}+i\gamma_{0,2})/2,~\chi_2=(\gamma_{0,3}+i\gamma_{0,4})/2$, where $\gamma_{0,j}$ is the Majorana edge mode localized at $\eta_j$~(see Appendix~\ref{appen:G2Mvort}). They satisfy $\chi_1|00\rangle=\chi_2|00\rangle=0,~|11\rangle=\chi^\dagger_1\chi^\dagger_2|00\rangle$. Therefore, they are orthogonal and have equal norm. In Sect.~\ref{sec:NAB} we will use Eqs.~\eqref{eq:g_1234}-\eqref{eq:basis0011} to study the non-Abelian statistics of the vortices.

\subsection{General $2M$ vortices case}\label{sec:Gen2Mvort}
We now study the general case with $2M$ vortices. The direct generalization of Eqs.~\eqref{eq:g_12} and \eqref{eq:g_1234} is 
\begin{equation}\label{eq:g_B}
g_B(z,z')=\frac{1}{4\pi}\frac{1}{z-z'}
\left[\sqrt{\frac{(z-\eta_{\bar{B}})!(z'-\eta_{B})!}{(z-\eta_B)!(z'-\eta_{\bar{B}})!}}+(z\leftrightarrow z')\right],
\end{equation}
where $B$ is an arbitrary subset of $S=\{1,2\ldots 2M\}$ with exactly $M$ elements, $|B|=M$, $\bar{B}\equiv S-B$, and $(z-\eta_{B})!$ is a collective symbol defined as
\begin{equation}
(z-\eta_B)!\equiv\prod_{j\in B}(z-\eta_j).
\end{equation}
There are $\binom{2M}{M}$ different choices of $B$, giving rise to $\binom{2M}{M}/2$ different ground states $|G_{[\alpha,g_{B}]}\rangle$~(notice that $g_B$ and $g_{\bar{B}}$ are always the same). However, these states are linearly dependent and they form an overcomplete basis of the ground state subspace. In Appendix~\ref{appen:2Mdim} we prove that the space spanned by all these states is actually $2^{M-1}$-dimensional.
This is consistent with the fact that with $2M$ vortices there are $2M$ independent Majorana edge modes $\gamma_{0,1}\ldots \gamma_{0,2M}$, which could be combined to $M$ Dirac fermion modes $\chi_{j}=(\gamma_{0,2j-1}+i\gamma_{0,2j})/2,~j=1\ldots M$, resulting in $2^{M}$ degenerate ground states, $2^{M-1}$ of which are in the even particle number sector.

We can construct an orthogonal basis for the~(even parity) ground state subspace as follows. Let $J$ denote a subset of $\{2j-1|1\leq j\leq M\}$ with an even number of elements. We try to construct a ground state $|G^J\rangle$ that is annihilated by $\chi_{j}$ if $(2j-1)\notin J$ and annihilated by $\chi_j^\dagger$ if $(2j-1)\in J$. In other words, we label the even fermion parity ground states by a set $J$ containing all the numbers $2j-1$ satisfying $\hat{n}_j|G^J\rangle=|G^J\rangle$, where $\hat{n}_j=\chi^\dagger_j\chi_j$~(hence $|J|$ is required to be even). It can be proved that~(see Appendix~\ref{sec:proof_g_J}) the ground state $|G^{J}\rangle=\sqrt{N_J}|G_{[\alpha,g^J]}\rangle$ with
\begin{eqnarray}\label{eq:g_J}
g^J(z,z')&=&\frac{1}{4\pi}\frac{1}{z-z'}\frac{1}{N_J}\sum_B\lambda^J_B
\left[\sqrt{\frac{(z-\eta_{\bar{B}})!(z'-\eta_{B})!}{(z-\eta_B)!(z'-\eta_{\bar{B}})!}}\right.\nonumber\\
&&\left.+(z\leftrightarrow z')\right]
\end{eqnarray}
satisfies this condition, where the summation is over all possible subsets $B$ of $S=\{1,2\ldots 2M\}$ such that exactly one of $\{2j-1,2j\}$ appears in $B$ for each $1\leq j\leq  M$~(there are $2^{M}$ different choices of $B$ in total). The coefficient $\lambda^J_B$ is defined as 
\begin{eqnarray}\label{eq:lambda^J_B}
\lambda^J_B=\prod_{i< j\in B}\sqrt{\eta_i-\eta_j}\sqrt{\eta_{\tilde{i}}-\eta_{\tilde{j}}}~(-1)^{|J\cap B|}
\end{eqnarray}%
where the notation $\tilde{m}$ exchanges $2j-1$ and $2j$ (e.g. $\tilde{1} = 2$, $\tilde{4} = 3$), and $N_J= \sum_B \lambda^J_B$  guarantees the correct residue of $g^J(z,z')$ at $z=z'$~[i.e. to satisfy Eq.~\eqref{eq:gzz} or the condition~(d) in Sect.~\ref{sec:2vort}]. There are $2^{M-1}$ different choices of $J$ in total, and the ground states $|G^J\rangle$ are orthogonal to each other~\footnote{We have also proved that in the cases of $2M=2,4,6$, all the states $|G^J\rangle$ have the same normalization factor, but we don't guarantee that this is also true for the general case.}. The ground states with odd fermion parity can be obtained by acting Majorana operators on $|G^J\rangle$, giving rise to $2^{M-1}$ linearly independent states~(for example, $\{\gamma_{0,j=1}|G^J\rangle,~\text{for}~\forall J\}$ is an orthogonal basis for the odd parity ground state subspace, see Appendix~\ref{sec:proof_g_J} for their explicit expressions).

\subsection{Non-Abelian statistics of the vortices}\label{sec:NAB}
The statistics of vortices in $p_x+ip_y$ superconductors is already established in a general setting in Ref.~[\onlinecite{Ivanov}], where the author searched for a unitary operator $\tau(T_i)$ that induces the braiding operation
\begin{equation}\label{eq:braiding_operation}
T_i: \begin{cases}
      \gamma_i\to\gamma_{i+1}, & \\
      \gamma_{i+1}\to-\gamma_{i}, & \\
     \gamma_j\to\gamma_{j}, & \text{for } j\neq i \text{ and } j\neq i+1.
    \end{cases}
\end{equation}
Namely, the author constructed a Hilbert space representation of the braid group in terms of the Majorana fermion operators. The explicit formula for this representation is
\begin{equation}\label{eq:rep_braid}
\tau(T_i)=\exp\left(\frac{\pi}{4}\gamma_{i+1}\gamma_i\right)=\frac{1}{\sqrt{2}}(1+\gamma_{i+1}\gamma_i),
\end{equation}
where $\gamma_i$ is the Majorana operator localized at vortex $\eta_i$. It can be easily checked that $\tau(T_i)$ given in Eq.~\eqref{eq:rep_braid} indeed induce the braiding operation Eq.~\eqref{eq:braiding_operation} in the sense $\tau(T_i)\gamma_j[\tau(T_i)]^{-1}=T_i(\gamma_j)$. Using the representation Eq.~\eqref{eq:rep_braid}, the author studied how ground states evolve under braiding, since the actions of $\gamma_i$ on the ground states are known. 

In this paper we use a different approach to study the non-Abelian statistics of the vortices. 
With the analytic wavefunctions Eq.~\eqref{eq:g_J} at hand, we can verify the non-Abelian statistics by explicitly doing the analytic continuation of the wavefunctions and calculating the Berry's matrix associated with the each braiding process. For simplicity we focus on the case of $2M=4$, where the wavefunctions are given in Eqs.~\eqref{eq:g_1234}-\eqref{eq:basis0011}. This approach is similar to the approach of  Ref.~[\onlinecite{Bonderson2011}], where the authors verified the non-Abelian statistics of the quasiholes
in $\nu=5/2$ FQH system by doing the explicit analytic continuation of the wavefunctions~(and calculating the associated Berry's matrix) under quasihole exchange. 

We denote by $\hat{B}_{j,j+1}$ the unitary evolution operator corresponding to a counterclockwise exchange of vortices $j$ and $j+1$. This is a quantum operator that acts on the space of ground states. One can decompose the operator ${\hat B}_{j,j+1}$ into a product of three terms: ${\hat B}_{j,j+1}=e^{i \phi} U^{(j,j+1)} {\tilde B}_{j,j+1}$, where  $e^{i\phi}$ is an overall Abelian phase factor~(including a dynamical phase, a geometrical phase and the Abelian part of the topological phase factor), which we will ignore henceforth, $U^{(j,j+1)}$ is the Berry's matrix, and ${\tilde B}_{j,j+1}$ accounts for the explicit evolution due to the analytic continuation of the wavefunctions defined in Eq.~\eqref{eq:basis0011}.  Specifically, $\tilde{B}_{j,j+1}$ accounts for the change of the wavefunctions when they are considered as functions of the $\{\eta_j\}$ and one counterclockwise exchanges vortices $\eta_j$ and $\eta_{j+1}$. For example, we have $\tilde{B}_{12}\sqrt{\eta_1-\eta_2}=i\sqrt{\eta_1-\eta_2}$, $\tilde{B}_{12}\sqrt{\eta_1-\eta_3}=\sqrt{\eta_2-\eta_3}$, etc. 

Hence the transformation of the ground state under adiabatically exchanging vortices $j$ and $j+1$ is
\begin{equation}\label{eq:combined_trans}
|G_\alpha\rangle\to \sum_\beta U^{(j,j+1)}_{\beta\alpha} \tilde{B}_{j,j+1}|G_\beta\rangle\equiv \hat{B}_{j,j+1}|G_\alpha\rangle,
\end{equation}
where $\alpha,\beta$ are collective labels of ground states. 
It has been proved in Ref.~[\onlinecite{Bonderson2011}] that if the basis wavefunctions $|G_\alpha\rangle$ satisfy the following two properties:\\ 
(1) $\langle G_\alpha|G_\beta\rangle=C \delta_{\alpha\beta}$, where $C$ is a constant independent of $\alpha,\beta$;\\ 
(2) All the basis wavefunctions $|G_\alpha\rangle$ are complex analytic~(can be multivalued) functions of all the vortex positions $\eta_j$, $j=1,2,\ldots, 2M$,\\
then the Berry's matrix is proportional to the identity matrix $U^{(j,j+1)}_{\beta\alpha}\propto \delta_{\beta\alpha}$. Since the above two conditions are satisfied by our wavefunctions in Eq.~(\ref{eq:basis0011})~(remember that wavefunctions are only defined at $z\neq \eta_j$ and vortices are well-separated $\eta_i\neq \eta_j$ for $\forall i,j$), $\hat{B}_{j,j+1}$ is equal to $\tilde{B}_{j,j+1}$ up to an overall Abelian phase factor. 

We first analyze the braiding of vortices $1,2$. Here we need to calculate $\tilde{B}_{12}|00\rangle$ and $\tilde{B}_{12}|11\rangle$. Notice that $\tilde{B}_{12}$ induces the following transformation rules
\begin{eqnarray}
|G_{13,24}\rangle&\leftrightarrow& |G_{14,23}\rangle,\nonumber\\
~\lambda_{00}&\leftrightarrow& \bar{\lambda}_{00},\nonumber\\
\sqrt{N_{00}}&\to&\sqrt{N_{00}},\nonumber\\
\sqrt{N_{11}}&\to& i\sqrt{N_{11}},
\end{eqnarray}  
leading to the evolution 
\begin{equation}\label{eq:B_12}
|00\rangle\to |00\rangle,~~|11\rangle\to i|11\rangle.
\end{equation}
Similarly, it can be checked that $\tilde{B}_{34}$ gives the same transformation.

More interesting things happen if we braid vortices $2,3$. We have
\begin{equation}\label{eq:B_23}
\tilde{B}_{23}|00\rangle=\sqrt{N'}(\lambda'|G_{12,34}\rangle+\bar{\lambda'}|G_{14,23}\rangle),
\end{equation}
where $N'=\sqrt{\eta_{12}\eta_{34}}+i\sqrt{\eta_{14}\eta_{23}}=\frac{i}{2}(\sqrt{N_{00}}-i\sqrt{N_{11}})^2$, and $\lambda'=\sqrt{\eta_{12}\eta_{34}}/N'$. 
Expanding $|G_{12,34}\rangle,|G_{14,23}\rangle$ in the basis of $|00\rangle,|11\rangle$ using Eqs.~(\ref{eq:G_1234}) and (\ref{eq:basis0011}) we get
\begin{equation}\label{eq:B_23_f}
\tilde{B}_{23}|00\rangle=\frac{e^{\frac{\pi}{4}i}}{\sqrt{2}}(|00\rangle-i|11\rangle).
\end{equation}
Similarly we have
\begin{equation}\label{eq:B_23_11}
\tilde{B}_{23}|11\rangle=\frac{e^{\frac{\pi}{4}i}}{\sqrt{2}}(|11\rangle-i|00\rangle).
\end{equation}
In summary we have the matrix representations of braid operators
\begin{eqnarray}
\hat{B}_{12}=\hat{B}_{34}=
\begin{bmatrix}
1&0\\
0&i
\end{bmatrix},~~~~
\hat{B}_{23}=\frac{e^{\frac{\pi}{4}i}}{\sqrt{2}}
\begin{bmatrix}
1&-i\\
-i&1
\end{bmatrix},
\end{eqnarray}
which exactly reproduce the results in Ref.~[\onlinecite{Ivanov}]~(which considered the universal properties for a $p_x+ip_y$ superconductor), up to an overall phase factor $e^{\pi i/4}$. [Note: The Abelian phase factor cannot be determined by Ivanov's approach since he was constructing a projective representation of the braid group. In our approach we have been ignoring an overall Abelian phase factor; to calculate this Abelian phase factor~(including a geometric part and a topological part), we need to calculate the normalization factor $C=\langle G_\alpha|G_\alpha\rangle$ as a function of the vortex positions $\eta_j$. This proves to be difficult, so we leave this as an open question.]

\section{Number-conserving Hamiltonians with exact projected mean-field states}\label{sec:NC}
As the mean-field picture of topological superconductors breaks particle number-conservation, there have been considerable theoretical effort to characterize Majorana zero modes in number-conserving, interacting systems\cite{Bos1,Bos2,Bos3,PZoller,RGK,SDiehl,NLang}, which is important in cold atom realizations and in understanding how much the mean-field description survives in interacting systems. Previous approaches are mostly focused on the 1D case, where special tools are available. In this section we show how to use the special point studied throughout this paper to construct 2D number-conserving Hamiltonians with exact projected mean-field states, following the general construction in Ref.~[\onlinecite{ZWang}]. Let's consider translation invariant geometries first. Recall from Sect.~\ref{sec:TIG} that the ground state $|G\rangle$ is annihilated by the operator $2\partial_{\bar{z}}\psi_z-\psi^\dagger_z$ for $\forall z\in S$. Therefore, defining $ \hat{A}_{zz'}\equiv \psi^\dagger_{z'}2\partial_{\bar{z}}\psi_z+\psi^\dagger_{z}2\partial_{\bar{z}'}\psi_{z'}$, we have
\begin{eqnarray}
 \hat{A}_{zz'}|G\rangle&\equiv& (\psi^\dagger_{z'}2\partial_{\bar{z}}\psi_z+\psi^\dagger_{z}2\partial_{\bar{z}'}\psi_{z'})|G\rangle\nonumber\\
 &=&[\psi^\dagger_{z'}(2\partial_{\bar{z}}\psi_z-\psi^\dagger_z)+\psi^\dagger_{z}(2\partial_{\bar{z}'}\psi_{z'}-\psi^\dagger_{z'})]|G\rangle\nonumber\\
  &=&0.
\end{eqnarray}
Therefore, $|G\rangle$ must be a ground state of the number-conserving Hamiltonian
\begin{equation}  \label{eq:Hparen_pos}
\hat{H}=\int_S W(z_1,z_2;z_3,z_4)A^\dagger_{z_1, z_2}A_{z_3,
z_4}\prod^4_{j=1}d^2z_j,
\end{equation}
where $W(z_1,z_2;z_3,z_4)$ is a positive semidefinite Hermitian matrix~\footnote{The matrix $W(z_1,z_2;z_3,z_4)$ has to satisfy some additional requirements to guarantee the uniqueness of the ground state in a definite particle number sector. These requirements are quite involved and technical. Here we only mention that for the Hamiltonian given in Eq.~\eqref{eq:H_pipsimple}, the projected mean-field ground state is the unique ground state in a definite particle number sector, with (anti-) periodic boundary conditions.}, in the sense that $\int f^*(z_1,z_2)W(z_1,z_2;z_3,z_4)f(z_3,z_4)\prod^4_{j=1}d^2z_j\geq 0$, for an arbitrary function $f(z_1,z_2)$.  For example, we can choose $W(z_1,z_2;z_3,z_4)=e^{-\lambda|z_1-z_3|}%
\delta^2(z_1-z_2)\delta^2(z_3-z_4)/4$ in Eq.~(\ref{eq:Hparen_pos}) where $\lambda>0$, 
after rearranging terms we get 
\begin{eqnarray}\label{eq:H_pipsimple}
\hat{H}&=&\int_S\nabla\psi^\dagger_{z}\cdot\nabla \psi_{z}d^2 z \\
&+&4\int_S e^{-\lambda|z-z'|}\psi^{\dagger}_{z}(\partial_{z^{\prime
}}\psi^\dagger_{z^{\prime }})\psi_{z^{\prime }}\partial_{\bar{z}}\psi_z d^2z
d^2z^{\prime }.  \notag
\end{eqnarray}

The ground state of $\hat{H}$~[for both Eqs.~\eqref{eq:Hparen_pos} and \eqref{eq:H_pipsimple}] with $2N$ particles~(for $--,-+,+-$ boundary conditions) is given by the projected mean-field state
\begin{eqnarray}
  |G_{2N}\rangle&=&\hat{P}_{2N}|G_{[g]}\rangle\nonumber\\
  &=&\frac{1}{N!}\left[\frac{1}{2}\int g(z,z')\psi^\dagger_z\psi^\dagger_{z'}d^2zd^2z'\right]^N|0\rangle,
\end{eqnarray}
and the ground state $|G_{2N+1}\rangle$ for $++$ periodic boundary condition could be obtained in a similar way.

In translation non-invariant geometries with walls or vortices, all the degenerate ground states $|G\rangle$ are annihilated by the operator $2\partial_{\bar{z}}\psi_z-2\partial_{\bar{z}}\ln\alpha~\psi_z-\alpha^2\psi^\dagger_z$ where $\alpha(z)$ is given in Eq.~\eqref{eq:alpha(x)_dbwall} for the wall case and in Eq.~\eqref{eq:alpha(z)_vort} for the vortices case. A second set of annihilators are the chiral edge modes with positive energy: $\gamma_{k,L},\gamma_{-k,R}$ for $k>0$ in the wall case and $\gamma_{j,n}$ for $j=1,\ldots,2M,~n>0$ in the vortices case, they all need to annihilate the ground states to minimize $\hat{K}'_{1}$ in Eqs.~\eqref{eq:K'_1diag} and Eq.~\eqref{eq:K'_1diag_vort}. From now on let us focus on the case with vortices for simplicity. Let $\gamma_{j,n}=\hat{u}_{j,n}+\hat{v}^\dagger_{j,n}$, where $\hat{u}_{j,n},\hat{v}^\dagger_{j,n}$ denote the creation and annihilation part of $\gamma_{j,n}$~[see Eq.~\eqref{def:gamma_n}], respectively. Three different sets of number-conserving operators are
\begin{eqnarray}\label{eq:NCanni_vort}
  \hat{A}^{(1)}_{zz'}&=& \alpha^2(z')\psi^\dagger_{z'}(2\partial_{\bar{z}}\psi_z-2\partial_{\bar{z}}\ln\alpha~\psi_z)+(z\leftrightarrow z')\nonumber\\
  \hat{A}^{(2)}_{zn}&=& \hat{v}^\dagger_n(2\partial_{\bar{z}}\psi_z-2\partial_{\bar{z}}\ln\alpha~\psi_z)-\alpha^2(z)\psi^\dagger_z\hat{u}_n\nonumber\\
 \hat{A}^{(3)}_{mn}&=&\hat{v}^\dagger_n\hat{u}_m+\hat{v}^\dagger_m\hat{u}_n,
\end{eqnarray}
all of which annihilate $|G\rangle$. The number-conserving Hamiltonian can in general be constructed as
\begin{eqnarray}  \label{eq:Hparen_gen}
\hat{H}=\sum_{i,j}H_{ij}\hat{A}^\dagger_i\hat{A}_j,
\end{eqnarray}
where we use $j$ to collectively label the annihilators in Eq.~\eqref{eq:NCanni_vort}, $H_{ij}$ is a positive semi-definite Hermitian matrix as before~\footnote{Other conditions on $H_{ij}$ may be desirable to impose, for example so that the Hamiltonian is short-ranged interacting, i.e. it should not couple operators separated far apart.}. The explicit form of $\hat{H}$ may be quite involved, but they are essentially the translation invariant version Eq.~\eqref{eq:Hparen_pos} with modifications at the boundary.

\section{Summary}\label{sec:concl}
We have shown how to obtain exact analytic expressions of the ground state wavefunctions in Read and Green's mean-field model of $p_x+ip_y$ superconductor at the special point $\mu=m\Delta^2/2$. By taking an appropriate ansatz wavefunction, the task of solving many-body Schr\"{o}dinger equation reduces to solving a first order partial differential equation of two complex variables subject to certain boundary conditions. Using this method, we obtained ground state wavefunctions in translation non-invariant geometries with walls and vortices, for a Hamiltonian with special modified boundary terms. As an application of these wavefunctions, we studied the non-Abelian statistics of the vortices by doing the explicit analytic continuation of these wavefunctions and calculating the associated Berry's matrix. The evolution matrices obtained this way exactly reproduce the result of an earlier paper~\cite{Ivanov} that was based on general arguments.

\acknowledgements
We thank Matthew Foster, Han Pu, Youjiang Xu, and Victor Gurarie for discussions. This work
was supported in part with funds from the Welch foundation, Grant No. C-1872.  K.R.A.H. thanks the Aspen Center for Physics, which is supported
by the National Science Foundation grant PHY-1066293,
for its hospitality while part of this work was performed.

\appendix
\section{Derivation of some identities}\label{app:identities}
\subsection{Proof of Eq.~\eqref{eq:deltafunction}}\label{app:identities_1}
Here we need to prove
$$
\frac{1}{\pi}\partial_{\bar{z}}\frac{1}{z-z'}=\delta^2(z-z').$$
First notice that when $z\neq z'$ where  $1/(z-z')$ is complex analytic, both sides vanish. Near $z= z'$, we have 
$$\int_S d^2 z \frac{1}{\pi}\partial_{\bar{z}}\frac{1}{z-z'}=-\frac{i}{2\pi}\oint_{\partial S} dz \frac{1}{z-z'}=1,$$
where $S$ is a circle region $|z-z'|\leq \delta$, and we used Green's integral formula 
$$
2\int_S\partial_{\bar{z}}F(z,\bar{z})d^2z=\int_S\partial_{\bar{z}}F(z,\bar{z})dz d\bar{z}=-i\oint_{\partial S}F(z,\bar{z})dz.$$
\subsection{Proof of Eq.~\eqref{eq:2DFintegral}}\label{app:identities_2}
Let $k=k_x+ik_y,\bar{k}=k_x-ik_y$. We have 
\begin{eqnarray}
\int \frac{e^{i \mathbf{k}\cdot \mathbf{r}}}{k}d^2k&=&\frac{2}{iz}\int \frac{1}{k}\partial_{\bar{k}}e^{\frac{i}{2}(\bar{k}z+k\bar{z})}d^2k \nonumber\\
&=&\frac{2}{iz}\int \left\{\partial_{\bar{k}}\left[\frac{1}{k}e^{\frac{i}{2}(\bar{k}z+k\bar{z})}\right]-\pi\delta^2(k)\right\}d^2k\nonumber\\
&=&\frac{2\pi i}{z}-\frac{1}{z}\lim_{K\to\infty}\oint_{|k|=K}\frac{1}{k}e^{\frac{i}{2}(\bar{k}z+k\bar{z})}dk\nonumber
\end{eqnarray}
The last term can be evaluated by the residue theorem:
\begin{eqnarray}
&&\oint_{|k|=K}\frac{1}{k}e^{\frac{i}{2}(\bar{k}z+k\bar{z})}dk\nonumber\\
&=&\oint_{|k|=K}\frac{1}{k}\sum_{n\geq 0}\frac{(-1)^n}{(2n)!2^{2n}}(\frac{K^2}{k}z+k\bar{z})^{2n}dk\nonumber\\
&=&2\pi i\sum_{n\geq 0}\frac{(-1)^n}{(n!)^2}(K|z|/2)^{2n}\nonumber\\
&=&2\pi i~ J_0(K|z|),
\end{eqnarray}
where $J_\nu(x)$ is the Bessel function of the first kind. Since $\lim_{x\to\infty}J_0(x)=0$, we have
\begin{eqnarray}
\int \frac{e^{i \mathbf{k}\cdot \mathbf{r}}}{k}\mathrm{d}^2k=\frac{2\pi i}{z}.
\end{eqnarray}

\subsection{Proof of Eq.~\eqref{eq:g+-}}\label{app:identities_3}
We first prove that the functions $g_{+-},g_{-+},g_{--}$ given in Eq.~\eqref{eq:g+-} are solutions to Eq.~\eqref{eq:gzz} with the corresponding boundary conditions. The Jacobi theta functions $\theta_j(z|i),j=1,2,3,4$ satisfy the quasiperiodicity relations
\begin{equation}
\theta_j(z+1|i)=\zeta_x \theta_j(z|i),~~\theta_j(z+i|i)=\zeta_y e^{\pi-2\pi i z}\theta_j(z|i),
\end{equation}
where the values of $\zeta_{x,y}$ for $\theta_j(z|i)$ are listed in Table~\ref{tab:jacobi}. From this, it is easy to check that $g_{+-},g_{-+},g_{--}$ satisfy their corresponding boundary conditions, e.g. $g_{+-}(z+L,z')=+g_{+-}(z,z'),g_{+-}(z+iL,z')=-g_{+-}(z,z')$.  
\begin{table}
\centering
\begin{tabular}{ |c|c|c| } 

 \hline
       &$\zeta_x$ & $\zeta_y$\\
\hline
 $\theta_1(z|i)$ & -1 & -1 \\ 
 $\theta_2(z|i)$ & -1 & +1 \\ 
 $\theta_3(z|i)$ & +1 & +1 \\ 
 $\theta_4(z|i)$ & +1 & -1 \\
 \hline
\end{tabular}
\caption{Quasiperiodicity of Jacobi theta functions $\theta_j(z|i)$. }\label{tab:jacobi}
\end{table}

We now proceed to prove that $g_{+-},g_{-+},g_{--}$ satisfy Eq.~\eqref{eq:gzz}. Take $g_{+-}(z,z')$ as an example. Since $\theta_2(z|i)$ is holomorphic and the only~(first order) zero points of $\theta_1(z|i)$ are $z=a+bi,~a,b\in \mathbb{Z}$, it follows that $2\partial_{\bar{z}}g_{+-}(z,z')=0$ except at $z=z'+a+bi,~a,b\in\mathbb{Z}$. Near $z=z'$, for an arbitrary analytic function $f(z)$ we have 
\begin{eqnarray}
&&\int_{|z|\leq \delta} f(z)2\partial_{\bar{z}}g_{+-}(z,z') d^2z\nonumber\\
&=&\int_{|z|\leq \delta} 2\partial_{\bar{z}}[f(z)g_{+-}(z,z') ]d^2z\nonumber\\
&=&-i\oint_{|z|=\delta} [f(z)g_{+-}(z,z') ]dz\nonumber\\
&=& f(z')
\end{eqnarray}
where in the last line we used the fact that the residue of $g(z,z')$ at $z=z'$ is $1/(2\pi)$. Therefore $g_{+-}(z,z')$ satisfies Eq.~\eqref{eq:gzz} at $z=z'$. From the periodicity of $g_{+-}(z,z')$ we know that $g_{+-}(z,z')$ satisfies Eq.~\eqref{eq:gzz} in the whole complex plane.

The uniqueness of the solution to Eq.~\eqref{eq:gzz} in a given (anti-)~periodic boundary condition can be proved similarly to the case of infinite plane, using Liouville's theorem. Therefore, the expressions in Eq.~\eqref{eq:Fourier} and Eq.~\eqref{eq:g+-} must be equivalent as they both solve Eq.~\eqref{eq:gzz}.

\section{Regularization of the model}\label{appen:norm}
The ground states constructed in equations like Eq.~\eqref{eq:G_TIG1} have the problem of being non-normalizable since the function $g(z,z')$ always has a pole at $z=z'$. 
The divergence of the normalization results from the unphysical choice of setting $\Delta(k)$ to a constant in Eq.~\eqref{eq:Kmf} rather than having $\Delta(k)\to 0$ for large $k$. 
We can expose the physical meaning as well as the appropriate regularization of this divergence by considering our model as the limit of a series of well-defined models with normalizable ground states. As an example, we consider the family of mean-field Hamiltonians $\hat{K}^{(a)}=\hat{K}^{(a)}_{\mathrm{bulk}}+\hat{K}_{\mathrm{bound}}$ where
\begin{eqnarray}
\hat{K}^{(a)}_{\mathrm{bulk}}&=&\int_S\left[2\partial_z\psi^\dagger_z-\int_S\delta_a^2(z-z')\psi_{z'}d^2z'\right]\nonumber\\
&\times&\left[2\partial_{\bar{z}} \psi_z-\int_S\delta_a^2(z-z'')\psi^\dagger_{z''}d^2z''\right] d^2 z
\end{eqnarray}
and 
\begin{equation}
\delta^2_a(z-z')=\frac{1}{2\pi a^2}\exp\left(-\frac{|z-z'|}{a}\right),
\end{equation}
with $a>0$ being a real constant. Since $\mathrm{lim}_{a\to 0}\delta^2_a(z-z')=\delta^2(z-z')$, the mean-field Hamiltonian $\hat{K}$ in Eq.~(\ref{eq:K_2Dpp}) can be considered as the $a\to 0$ limit of $\hat{K}^{(a)}$. It is easy to check that the ground states of $K^{(a)}$ could still be constructed by the exponential form of Eq.~(\ref{eq:Gg}) in the main text with $g(z,z')$ satisfying
\begin{equation}\label{eq:gzz_a}
2\partial_{\bar{z}} g(z,z')=\delta^2_a(z-z').
\end{equation}
With the sharp delta function in Eq.~(\ref{eq:gzz}) replaced by a well-defined function $\delta^2_a(z-z')$, the pole of $g(z,z')$ is removed. For example, in regions far from edges and vortices, Eq.~(\ref{eq:gzz_a}) has solution
\begin{equation}
g_a(z,z')=\frac{1}{2\pi(z-z')}\left[1-\left(1+\frac{|z-z'|}{a}\right)\exp\left(-\frac{|z-z'|}{a}\right)\right].
\end{equation}
In the limit $a\to 0$, $g_a(z,z')$ gives back Eq.~\eqref{eq:G_TIG1} in the main text, but for any nonzero $a$, $g_a(z,z')$ is free of poles and in particular we have $g_a(z,z)=0$, and the ground state is normalizable in a finite system. This justifies our use of non-normalizable ground states.

\section{The boundary Hamiltonian $\hat{K}'$ and  chiral edge modes $\gamma_{k,a}$}\label{appen:K'}
In this section we present the mathematical details about how to modify the Hamiltonian near the boundary so that the system has zero-energy ground states. We will show the definition of functions $\Delta(z,z')$ and $\mu(z,z')$ in Eq.~\eqref{eq:K'_1} and give the derivations from Eq.~\eqref{eq:K'_1diag} to Eq.~\eqref{eq:bc_vort}.

\subsection{The Double Wall Geometry}\label{appen:K'_dbwall}
For simplicity we focus on the left boundary, i.e. we only consider $\gamma_{k,L}$~(abbreviated as $\gamma_k$) and $\Delta(z,z'),\mu(z,z')$ near the left boundary, and take the right boundary to be sufficiently far away.

As we mentioned in the main text, the functions $\Delta(z,z'),\mu(z,z')$ in Eq.~\eqref{eq:K'_1} are chosen so that $\hat{K}'_1$
commutes with $2\partial_{\bar{z}}\psi_{z}-2\partial_{\bar{z}}\ln\alpha\psi_{z}-\alpha^2\psi_{z}^{\dagger }$ for $\forall z\in S$. This leads to differential equations for $\Delta(z,z'),\mu(z,z')$
\begin{eqnarray}\label{eq:diff_Deltamu}
2\partial_{\bar{z}}\left[\frac{\mu(z,z')}{\alpha(z)}\right]+\alpha(z)\Delta^*(z,z')=0,\nonumber\\
2\partial_{z}\left[\frac{\Delta^*(z,z')}{\alpha(z)}\right]+\alpha(z)\mu(z,z')=0.
\end{eqnarray}
Without loss of generality, we can require the function $\Delta(z,z')$  to be anti-symmetric $\Delta(z,z')=-\Delta(z',z)$ and $\mu(z,z')$ to be Hermitian $\mu^*(z,z')=\mu(z',z)$, and in the double wall geometry they are both required to be translation invariant in the $y$-direction~(depend only on $y-y'$).  
To find a nontrivial solution to Eq.~\eqref{eq:diff_Deltamu} we try the ansatz solution
\begin{eqnarray}\label{eq:deltamu_sol}
\Delta(z,z')&=& \int^{+\infty}_{-\infty}k v_{-k}(x)v_k(x') e^{-ik(y-y')}dk ,\nonumber\\
\mu(z,z')&=&\int^{+\infty}_{-\infty}k v_{-k}(x)u_k(x')e^{-ik(y-y')}dk,
\end{eqnarray}
where the functions $u_k(x),v_k(x)$ satisfy $u_{-k}(x)=v^*_k(x)$ and
\begin{eqnarray}\label{eq:diff_uv}
(\partial_x-k)\left[\frac{v_k(x)}{\alpha(x)}\right]=\alpha(x)u_k(x),\nonumber\\
(\partial_x+k)\left[\frac{u_k(x)}{\alpha(x)}\right]=\alpha(x)v_k(x),
\end{eqnarray}
and the normalization condition
\begin{equation}\label{eq:uvnorm}
\int^\infty_0 [|u_k(x)|^2+|v_k(x)|^2]dx=1.
\end{equation}
Though we could analytically solve Eq.~\eqref{eq:diff_uv} only in the large-$\lambda$ limit~(which we'll present in a moment), we can prove that there is a unique solution localized at the boundary. With these solutions, Eq.~\eqref{eq:deltamu_sol} completes the definition of $\hat{K}'_1$. Inserting Eq.~\eqref{eq:deltamu_sol} into Eq.~\eqref{eq:K'_1}, we get Eq.~\eqref{eq:K'_1diag}, where 
\begin{equation}\label{def:gamma_k}
\gamma_k=\int_S e^{ik y}[u_k(x)\psi_z+v_k(x)\psi^\dagger_z]d^2z,
\end{equation}
and, in light of Eq.~\eqref{eq:diff_uv} one can show that $\{\gamma_k,2\partial_{\bar{z}}\psi_{z}-2\partial_{\bar{z}}\ln\alpha\psi_{z}-\alpha^2\psi_{z}^{\dagger }\}=0$. The normalization condition then gives rise to $\{\gamma_k,\gamma_{k'}\}=\delta(k+k')$. This confirms Eq.~\eqref{eq:b_kcommute}.

Let us now investigate how $\gamma_k$ acts on the ground state $|G\rangle$ in Eq.~\eqref{eq:Ggalpha}. Denoting $u_k(z)=e^{iky}u_k(x),v_k(z)=e^{iky}v_k(x)$, we have
\begin{eqnarray}\label{eq:gamma_kG}
\gamma_k|G\rangle&=&\int_S [u_k(z)\psi_z+v_k(z)\psi^\dagger_z]d^2z|G\rangle\nonumber\\
&=&\int_S \left\{\frac{1}{\alpha}2\partial_{\bar{z}}\left[\frac{v_k(z)}{\alpha}\right]\psi_z+v_k(z)\psi^\dagger_z\right\}d^2z|G\rangle\nonumber\\
&=& \int_S \left[\psi^\dagger_z-\frac{1}{\alpha}2\partial_{\bar{z}}\left(\frac{\psi_z}{\alpha}\right)\right]v_k(z)d^2z|G\rangle\nonumber\\
&-&i\oint_{\partial S}\frac{v_k(z)\psi_z}{\alpha^2}dz|G\rangle,
\end{eqnarray}
where in the second line we used Eq.~\eqref{eq:diff_uv} and in the third line we integrated by parts. The term in the third line of Eq.~\eqref{eq:gamma_kG} is zero since $|G\rangle$ is annihilated by $\alpha\psi^\dagger_z-2\partial_{\bar{z}}(\psi_z/\alpha)$. Therefore
\begin{eqnarray}\label{eq:gamma_kG_2}
\gamma_k|G\rangle&=&-i\oint_{\partial S}\frac{v_k(z)\psi_z}{\alpha^2(z)}dz|G\rangle\nonumber\\
&=&-i\oint_{\partial S}\frac{v_k(z)}{\alpha(z)}dz  \int_{S}g(z,z^{\prime })\alpha(z')\psi_{z^{\prime }}^{\dagger }d^{2}z^{\prime }|G\rangle\nonumber\\
&\equiv&\int_{z'\in S}w_k(z')\alpha(z')\psi_{z^{\prime }}^{\dagger }d^{2}z^{\prime }|G\rangle,
\end{eqnarray}
where in the second line we used Eq.~\eqref{eq:Ggalpha} and defined
\begin{equation}\label{eq:w_k(z)}
w_k(z')=-i\oint_{\partial S}\frac{v_k(z)}{\alpha(z)} g(z,z^{\prime })dz.
\end{equation}
Since the ground state $|G\rangle$ should be annihilated by $\gamma_k$ for $\forall k>0$, we need to require $w_k(z')\equiv 0$ for $\forall k>0$, which is equivalent to Eq.~\eqref{eq:bc_dbwall-2} in the main text. We now explicitly check that the function $g(z,z')=1/2\pi(z-z')$ satisfies the condition Eq.~\eqref{eq:bc_dbwall-2} for $k>0$. By completing the contour integration and using the residue theorem, we have
\begin{eqnarray}\label{eq:contintformula}
\int^\infty_{-\infty}\frac{e^{ik y}dy}{iy-z'}=\begin{cases}
      0, & k>0, \\
      \pi, & k=0,\\
      2\pi e^{k z'} & k<0.
    \end{cases}
\end{eqnarray}
Applying Eq.~\eqref{eq:contintformula} to Eqs.~\eqref{eq:w_k(z)} and \eqref{eq:gamma_kG_2}, we get $\gamma_k|G\rangle=0$ for $k>0$,
\begin{eqnarray}\label{eq:gamma_k0G_2}
\gamma_{k=0}|G\rangle=\frac{v_{k=0}(0)}{2\alpha(0)}\int_{z'\in S}\alpha(z')\psi_{z^{\prime }}^{\dagger }d^{2}z^{\prime }|G\rangle,
\end{eqnarray}
which verifies Eq.~\eqref{eq:G_dbwall} and 
\begin{eqnarray}\label{eq:gamma_k>0G_2}
\gamma^\dagger_{k}|G\rangle=\frac{v_{-k}(0)}{\alpha(0)}\int_{z'\in S}\alpha(z')e^{-kz'}\psi_{z^{\prime }}^{\dagger }d^{2}z^{\prime }|G\rangle,
\end{eqnarray}
from which we can calculate $\gamma_k\gamma_{k'}|G\rangle$:
\begin{eqnarray}
&&\gamma^\dagger_k\gamma^\dagger_{k'}|G\rangle
=\frac{v_{-k'}(0)v_{-k}(0)}{\alpha(0)^2}\times\nonumber\\
&&\int_S \alpha(z)\alpha(z')e^{-kz-k'z'}\psi^\dagger_{z}\psi^\dagger_{z'}d^2z d^2z'|G\rangle,
\end{eqnarray}
thus, setting $c_k=\alpha(0)/v_{-k}(0)$, we have
\begin{eqnarray}
&&(1+\xi c_kc_k' \gamma^\dagger_k\gamma^\dagger_{k'})|G\rangle=\nonumber\\
&&\exp\left[-\xi\int\alpha(z)\alpha(z')e^{-kz-k'z'}\psi^\dagger_z\psi^\dagger_{z'}d^2zd^2z'\right]|G\rangle,\nonumber
\end{eqnarray}
which confirms Eqs.~\eqref{eq:g_kk'} and \eqref{eq:G_kk'} in the main text.

We finally mention that the differential equations Eq.~\eqref{eq:diff_uv} can be analytically solved in the large-$\lambda$ limit, in which $\alpha(x)=1-e^{-\lambda x}$ becomes a step function. The solution is 
\begin{eqnarray}\label{sol:uv}
u_k(x)&=&\frac{i}{\sqrt{2\pi}}\sqrt{\sqrt{k^2+1}+k} ~\alpha(x)e^{-\sqrt{k^2+1}x},\nonumber\\
v_k(x)&=&\frac{-i}{\sqrt{2\pi}}\sqrt{\sqrt{k^2+1}-k} ~\alpha(x)e^{-\sqrt{k^2+1}x},
\end{eqnarray}
and $\Delta(z,z'),\mu(z,z')$ have simpler expresions 
\begin{eqnarray}\label{sol:Deltamu}
  \Delta(z,z')&=&\int^\infty_{-\infty}\frac{k}{2\pi}\nonumber\\
                 &&\exp[-\sqrt{k^2+1}(x+x')+ik(y-y')]dk,\nonumber\\
\mu(z,z')&=&(\partial_x-i\partial_y)\Delta(z,z').
\end{eqnarray}
All the important results in this paper are, however, independent of this limit.

\subsection{The Vortices Geometry}\label{appen:K'_vort}
Since the vortices are far apart, we can treat them independently when discussing the well-localized chiral edge modes. For simplicity, in this section we focus on any one of them, e.g. the $j$-th vortex. For simplicity, we can set $\eta_j$ at the origin $\eta_j=0$. We abbreviate $\gamma_{n,j}$ as $\gamma_n$ and only study $\Delta(z,z'),\mu(z,z')$ near the $j$-th vortex.

Our discussion of the vortex chiral edge modes is essentially parallel to that in the double wall geometry. 
In order that $\hat{K}'_1$ commutes with $2\partial_{\bar{z}}\psi_{z}-2\partial_{\bar{z}}\ln\alpha\psi_{z}-\alpha^2\psi_{z}^{\dagger }$ for $\forall z\in S$, the  functions $\Delta(z,z'),\mu(z,z')$ should again satisfy differential equations Eq.~\eqref{eq:diff_Deltamu},
and we again we require $\Delta(z,z')$  to be anti-symmetric $\Delta(z,z')=-\Delta(z',z)$ and $\mu(z,z')$ to be Hermitian $\mu^*(z,z')=\mu(z',z)$. In this case they are both required to  be anti-periodic around the vortex $\Delta(z e^{2\pi i},z')=-\Delta(z,z'),\mu(z e^{2\pi i},z')=-\mu(z,z')$.  
To find a nontrivial solution to Eq.~\eqref{eq:diff_Deltamu} we try the ansatz solution~(where $z=r e^{i\theta},z'=r' e^{i\theta'}$)
\begin{eqnarray}\label{eq:deltamu_sol_vort}
\Delta(z,z')&=& \sum^{+\infty}_{n=1}n v_{-n}(r)v_n(r') e^{in(\theta'-\theta)}e^{-i(\theta'+\theta)/2} ,\nonumber\\
\mu(z,z')&=&\sum^{+\infty}_{n=1}n v_{-n}(r)u_n(r')e^{i(n+1/2)(\theta'-\theta)},
\end{eqnarray}
where the functions $u_n(r),v_n(r)$ satisfy $u_{-n}(r)=v^*_n(r)$ and
\begin{eqnarray}\label{eq:diff_uv_vort}
\left(\partial_r-\frac{n-1/2}{r}\right)\left[\frac{v_n(r)}{\alpha(r)}\right]=\alpha(r)u_n(r),\nonumber\\
\left(\partial_r+\frac{n+1/2}{r}\right)\left[\frac{u_n(r)}{\alpha(r)}\right]=\alpha(r)v_n(r),
\end{eqnarray}
and the normalization condition
\begin{equation}\label{eq:uvnorm_vort}
\int^\infty_0 [|u_n(r)|^2+|v_n(r)|^2]rdr=1.
\end{equation}
Similarly to the double-wall case, by inserting Eq.~\eqref{eq:deltamu_sol_vort} into Eq.~\eqref{eq:K'_1} we get Eq.~\eqref{eq:K'_1diag_vort} where
\begin{equation}\label{def:gamma_n}
\gamma_n=\int_S [u_n(z)\psi_z+v_n(z)\psi^\dagger_z]d^2z,
\end{equation}
where $u_n(z)=u_n(r)e^{i(n+1/2)\theta},v_n(z)=v_n(r)e^{i(n-1/2)\theta}$. Eq.~\eqref{eq:diff_uv_vort} implies that $\{\gamma_k,2\partial_{\bar{z}}\psi_{z}-2\partial_{\bar{z}}\ln\alpha\psi_{z}-\alpha^2\psi_{z}^{\dagger }\}=0$, and the normalization condition gives $\{\gamma_n,\gamma_{n'}\}=\delta_{n,-n'}$. This confirms Eq.~\eqref{eq:b_kcommute_vort}.

We now investigate how $\gamma_n$ acts on the ground states $|G\rangle$. The analog of Eq.~\eqref{eq:gamma_kG_2} in the vortex case is \footnote{Notice that the analog of Eq.~\eqref{eq:w_k(z)} changes sign on the right hand side because in this case the superconducting region $S$ lies outside of the counterclockwise contour.}
\begin{eqnarray}\label{eq:gamma_nG}
\gamma_n|G\rangle=\int_{z'\in S}w_n(z')\alpha(z')\psi_{z^{\prime }}^{\dagger }d^{2}z^{\prime }|G\rangle,
\end{eqnarray}
where 
\begin{equation}\label{eq:w_n(z)}
w_n(z')=i\oint_{|z-\eta_j|=r_0}\frac{v_{n}(z)}{\alpha(z)} g(z,z^{\prime })dz.
\end{equation}
 Since the ground state $|G\rangle$ should be annihilated by $\gamma_n$ for $\forall n>0$, we need to require $w_n(z')\equiv 0$ for $\forall n>0$:
\begin{eqnarray}\label{eq:bc_vort_1}
w_n(z')&=&i\oint_{|z-\eta_j|=r_0}\frac{v_n(z)}{\alpha(z)} g(z,z^{\prime })dz\nonumber\\
&=&i\frac{v_n(r_0)}{\alpha(r_0) r^{n-1/2}_0}\oint_{|z-\eta_j|=r_0}(z-\eta_j)^{n-1/2} g(z,z^{\prime })dz\nonumber\\
&=&0,
\end{eqnarray}
where we have used the fact that $\alpha(z)$ only depends on $|z-\eta_j|$ near the $j$-th vortex. Eq.~\eqref{eq:bc_vort_1} is equivalent to Eq.~\eqref{eq:bc_vort} in the main text. Then, Eqs.~\eqref{eq:G_o_2vort}-\eqref{eq:G_mn} can be directly verified using Eqs.~\eqref{eq:gamma_nG}-\eqref{eq:w_n(z)} and the contour integral formula~(for $|z'|>r_0$)
\begin{eqnarray}\label{eq:contintformula_2}
\frac{1}{2\pi i}\oint_{|z|=r_0}\frac{z^{n-1/2}}{z-z'}\left(\sqrt{\frac{z}{z'}}+\sqrt{\frac{z'}{z}}\right)dz\nonumber\\
=\begin{cases}
      0, & n\geq 0, \\
      -\frac{1}{\sqrt{z'}}, & n=0,\\
     -2(z')^{n-1/2} & n\leq -1.
    \end{cases}
\end{eqnarray}

Again, the differential equations Eq.~\eqref{eq:diff_uv_vort} can be analytically solved in the large-$\lambda$ limit~(up to an overall normalization factor):
\begin{eqnarray}\label{sol:uv}
u_n(r)&=&\alpha(r)H^{(1)}_{n+1/2}(ir),\nonumber\\
v_n(r)&=&\alpha(r)iH^{(1)}_{n-1/2}(ir),
\end{eqnarray}
where $H^{(1)}_{n}(x)$ is the Hankel function of the first kind. Equivalently, $u_n(z),v_n(z)$ can be represented as~(up to normalization)
\begin{eqnarray}
u_0(z)&=&-i \frac{e^{-|z-\eta_j|}}{\sqrt{\bar{z}-\bar{\eta}_j}},~~v_0(z)=i \frac{e^{-|z-\eta_j|}}{\sqrt{z-\eta_j}},\nonumber\\
u_n(z)&=&\partial_{\bar{z}}^n u_0(z),~v_n(z)=\partial_{\bar{z}}^n v_0(z),\nonumber\\
u_{-n}(z)&=&\partial_{z}^n u_0(z),~v_{-n}(z)=\partial_{z}^n v_0(z).
\end{eqnarray}

\section{Ground states with $2M$ vortices}\label{appen:G2Mvort}
In this section we prove some important statements and equations in Sect.~\ref{sec:vortices}. In Sect.~\ref{appen:2Mdim} we prove that the space of even fermion parity ground state wavefunctions corresponding to the solutions of Eqs.~\eqref{eq:gzz} and \eqref{eq:bc_vort} in $2M$ vortices geometry is $2^{M-1}$-dimensional.
In Sect.~\ref{sec:proof_g_J} we prove that Eq.~\eqref{eq:g_J} gives an orthogonal basis for this $2^{M-1}$-dimensional ground state subspace.
\subsection{Dimension of ground state subspace with $2M$ vortices}\label{appen:2Mdim}
In this subsection we prove that the dimension of the ground state subspace~(with even fermion parity) in $2M$ vortices geometry is $2^{M-1}$. A similar problem was encountered in Ref.~[\onlinecite{Nayak1996}] where the authors proved that the space of Moore-Read Pfaffian states with $2M$ quasiholes is $2^{M-1}$ dimensional. Despite the similarity, our proof is much simpler than theirs since in our case we are dealing with the exponentiation of $g(z,z')$~[see Eq.~\eqref{eq:Ggalpha}] rather than the Pfaffian of $g(z,z')$.

We follow our notations in Sect.~\ref{sec:Gen2Mvort}. As we have seen in the main text, the problem of finding all the degenerate ground states~(with even fermion parity) reduces to finding all possible solutions to the differential equation Eq.~\eqref{eq:gzz} subject to boundary condition Eq.~\eqref{eq:bc_vort}. As we already know that Eq.~\eqref{eq:g_B} is a solution, similar to Eq.~\eqref{eq:dif_g_12}, we can consider the difference between a general solution $g(z,z')$ and $g_B(z,z')$
\begin{equation}\label{eq:dif_gg_B}
g(z,z')=g_{B}(z,z')+\frac{f(z,z')}{\sqrt{(z-\eta_S)!(z'-\eta_S)!}}
\end{equation}
where $f(z,z')=-f(z',z)$. Similar to the discussion in Sect.~\ref{sec:2vort},  Eq.~(\ref{eq:gzz}) along with Eq.~\eqref{eq:bc_vort}~[or the condition~(d) in Sect.~\ref{sec:2vort}] requires that $f(z,z')$ is an analytic function of $z$ in the entire complex plane~(free of poles) except at $z=\infty$.  The asymptotic behavior of $g(z,z')$ for large $z$~[condition~(b) in Sect.~\ref{sec:2vort}] requires that $f(z,z')$ grows at most as fast as $z^{M-1}$ as $z\to \infty$~(for $z'$ fixed), since $\sqrt{(z-\eta_S)!}\sim z^{M}$ as $z\to \infty$. In summary, $f(z,z')$ is an anti-symmetric polynomial of $z,z'$ of degree~(in $z$) at most $M-1$. We expand $f(z,z')$ as
\begin{equation}
f(z,z')=\sum_{0\leq m<n\leq M-1}C_{mn}(z^mz'^n-z^nz'^m),
\end{equation}
where $C_{mn}=-C_{nm}$ are complex coefficients. Inserting Eq.~\eqref{eq:dif_gg_B} into Eq.~\eqref{eq:Ggalpha} we obtain the relation between the corresponding ground states $|G_g\rangle\equiv|G_{[\alpha,g]}\rangle$ and $|G_{g_B}\rangle\equiv|G_{[\alpha,g_B]}\rangle$:
\begin{eqnarray}\label{eq:rel_gg_B}
  |G_g\rangle&=&\exp\left[\frac{1}{2}\int\frac{f(z,z')\alpha(z)\alpha(z')}{\sqrt{(z-\eta_S)!(z'-\eta_S)!}}\right.\nonumber\\
                 &&\ \ \left.\psi^\dagger_z\psi^\dagger_{z'}d^2zd^2z'\right]|G_{g_B}\rangle\nonumber\\
&=&\prod_{0\leq m<n\leq M-1}\exp\left(C_{mn}\phi^\dagger_m\phi^\dagger_{n}\right)|G_{g_B}\rangle\nonumber\\
&=&\prod_{0\leq m<n\leq M-1}\left(1+C_{mn}\phi^\dagger_m\phi^\dagger_{n}\right)|G_{g_B}\rangle,
\end{eqnarray}
where we have defined linearly independent single fermion modes
\begin{equation}
\phi^\dagger_{n}=\int\frac{z^n}{\sqrt{(z-\eta_S)!}}\alpha(z)\psi^\dagger_zd^2z
\end{equation}
for $1\leq n\leq M-1$. The last line of Eq.~\eqref{eq:rel_gg_B} indicates that a general ground state $|G_g\rangle$ can always be represented as a linear superposition of the following basis states
\begin{equation}\label{eq:basis2M}
\left\{\prod^{2l}_{j=1}\phi^\dagger_{m_j}|G_{g_B}\rangle  \right\},
\end{equation}
where $l\in\mathbb{Z},~0\leq 2l\leq M,~0\leq m_1<\ldots<m_{2l}\leq M-1$. The basis states in Eq.~\eqref{eq:basis2M} are linearly independent, since otherwise we would have
\begin{equation}\label{eq:linear_dep}
\sum_{l\geq l_0}\left(\sum_{m_1,\ldots,m_{2l}} c^{(2l)}_{\{m\}}\prod^{2l}_{j=1}\phi^\dagger_{m_j}\right)|G_{g_B}\rangle=0,
\end{equation}
where we can assume $c^{(2l_0)}_{\{m\}}\neq 0$ for at least a set $\{m_1,\ldots,m_{2l}\}$ without loss of generality. Now insert the expansion $|G_{g_B}\rangle=\sum_{l\geq 0}\hat{P}_{2l}|G_{g_B}\rangle$~(where $\hat{P}_{2l}$ is the particle number projection operator) into Eq.~\eqref{eq:linear_dep} and consider the state with minimal number of particles on the left hand side~(remember that $\hat{P}_0|G_{g_B}\rangle=|0\rangle$), we get
\begin{equation}
  \left(\sum_{m_1,\ldots,m_{2l}} c^{(2l_0)}_{\{m\}}\prod^{2l_0}_{j=1}\phi^\dagger_{m_j}\right)|0\rangle=0,
\end{equation}
which is impossible since $\{\phi^\dagger_{m_j}\}$ are linearly independent fermion modes.
The total number of the basis states in Eq.~\eqref{eq:basis2M} is 
\begin{equation}\label{eq:sumC_2M}
\sum_{2l\leq M} {{M}\choose{2l}}=2^{M-1},
\end{equation}
meaning that the ground state subspace with even fermion parity is $2^{M-1}$ dimensional. 

We end this section by obtaining a useful identity that is valid for $M\leq 3$~(no more than six vortices), which can be applied to obtain Eqs.~\eqref{eq:G_1234} and \eqref{eq:basis0011} in Sect.~\ref{sec:4vort}. Define 
\begin{equation}
\hat{g}=\frac{1}{2}\int_{z,z'\in S} g(z,z')\alpha(z)\alpha(z')\psi^\dagger_z\psi^\dagger_{z'}d^2zd^2z',
\end{equation}
for an arbitrary solution $g(z,z')$ of Eqs.~\eqref{eq:gzz} and \eqref{eq:bc_vort}. Then we have the identity
\begin{equation}\label{eq:identityM3}
\exp\left(\sum_j \lambda_j \hat{g}_j\right)=\sum_j\lambda_j \exp(\hat{g}_j),
\end{equation}
where $g_j(z,z')$ are solutions to Eqs.~\eqref{eq:gzz} and \eqref{eq:bc_vort} and $\lambda_j$ are coefficients satisfying $\sum_j\lambda_j=1$.
The proof begins by noticing that when $M\leq 3$ we have
\begin{equation}
(\hat{g}_i-\hat{g}_j)^2=0,
\end{equation}
which is true since $\hat{g}_i-\hat{g}_j=C_{01}\phi^\dagger_0\phi^\dagger_1+C_{02}\phi^\dagger_0\phi^\dagger_2+C_{12}\phi^\dagger_1\phi^\dagger_2$ for $M\leq 3$ and $\phi^{\dagger 2}_j\equiv 0$~($C_{01},C_{02},C_{12}$ are unimportant coefficients). Therefore, we have~(notice that $\hat{g}_i,\hat{g}_j$ mutually commute)
\begin{eqnarray}
\exp\left(\sum_j \lambda_j \hat{g}_j\right)&=&e^{\hat{g}_1}\exp\left[\sum_j \lambda_j (\hat{g}_j-\hat{g}_1)\right]\nonumber\\
&=&e^{\hat{g}_1}\left[1+\sum_j \lambda_j (\hat{g}_j-\hat{g}_1)\right],\nonumber\\
\end{eqnarray}
while on the other hand we have
\begin{eqnarray}
\sum_j\lambda_j \exp(\hat{g}_j)&=&\sum_j\lambda_j \exp(\hat{g}_1+\hat{g}_j-\hat{g}_1)\nonumber\\
&=&\sum_j\lambda_j \exp(\hat{g}_1)(1+\hat{g}_j-\hat{g}_1)\nonumber\\
&=&e^{\hat{g}_1}\left[1+\sum_j \lambda_j (\hat{g}_j-\hat{g}_1)\right].
\end{eqnarray}
Combining the two equations above we prove Eq.~\eqref{eq:identityM3}.

We now proceed to prove Eq.~\eqref{eq:G_1234} and Eq.~\eqref{eq:basis0011} with the help of Eq.~\eqref{eq:identityM3}. Starting from the identity
\begin{equation}\label{eq:eta_1234_rel}
\eta_{12}\eta_{34}=\eta_{13}\eta_{24}-\eta_{14}\eta_{23},
\end{equation}
one can show that
\begin{equation}\label{eq:g_1234_rel}
\eta_{12}\eta_{34}~g_{12,34}=\eta_{13}\eta_{24}~g_{13,24}-\eta_{14}\eta_{23}~g_{14,23},
\end{equation}
therefore
\begin{eqnarray}\label{eq:G_1234_rel}
|G_{12,34}\rangle&=&\exp(\hat{g}_{12,34})\nonumber\\
                 &=&\exp\left(\frac{\eta_{13}\eta_{24}}{\eta_{12}\eta_{34}}\hat{g}_{13,24}-\frac{\eta_{14}\eta_{23}}{\eta_{12}\eta_{34}}\hat{g}_{14,23}\right)\nonumber\\
  &=&\frac{\eta_{13}\eta_{24}}{\eta_{12}\eta_{34}}\exp(\hat{g}_{13,24})-\frac{\eta_{14}\eta_{23}}{\eta_{12}\eta_{34}}\exp(\hat{g}_{14,23})\nonumber\\
  &=&\frac{\eta_{13}\eta_{24}}{\eta_{12}\eta_{34}}|G_{13,24}\rangle-\frac{\eta_{14}\eta_{23}}{\eta_{12}\eta_{34}}|G_{14,23}\rangle,
\end{eqnarray}
where in the third line we used Eq.~\eqref{eq:identityM3}. This proves Eq.~\eqref{eq:G_1234}. The state $|00\rangle$ in Eq.~\eqref{eq:basis0011} is constructed from Eq.~\eqref{eq:g_J} for the case $2M=4$ and $J=\emptyset$
\begin{eqnarray}
  |00\rangle&=&\sqrt{N_{J}}\exp(\hat{g}^{J})\nonumber\\
            &=&\sqrt{N_{00}}\exp(\lambda_{00}\hat{g}_{13,24}+\bar{\lambda}_{00}\hat{g}_{14,23})\nonumber\\
  &=&\sqrt{N_{00}}(\lambda_{00}|G_{13,24}\rangle+\bar{\lambda}_{00}|G_{14,23}\rangle),
\end{eqnarray}
and similarly for $|11\rangle$~(with $J=\{1,3\}$). Thus the states $|00\rangle,|11\rangle$ form the occupation number basis of the nonlocal fermions $\chi_1,\chi_2$~(see next section for the proof of the general case).

\subsection{Proof of Eq.~\eqref{eq:g_J}}\label{sec:proof_g_J}
In this subsection we prove that the ground states $|G_J\rangle=\sqrt{N_J}|G_{[\alpha,g^J]}\rangle$ with $g^J(z,z')$ defined in Eq.~\eqref{eq:g_J} form an orthogonal basis of the $2^{M-1}$-dimensional ground state subspace. We begin by studying the action of Majorana operators on $|G^J\rangle$, using Eqs.~\eqref{eq:gamma_nG} and \eqref{eq:w_n(z)}:

\begin{widetext}
\begin{eqnarray}\label{eq:gamma_jG^J}
\gamma_{0,j}|G^J\rangle&=&\int_{z'\in S}ic_0\oint_{|z-\eta_j|=r_0}\frac{g^J(z,z')}{\sqrt{z-\eta_j}}dz ~\alpha(z')\psi^\dagger_{z'}d^2z'|G^J\rangle\\
&=&-\frac{c_0}{N_J}\int_{z'\in S}\sum_{B\ni j}\lambda_B^J\frac{1}{\eta_j-z'}\sqrt{\frac{(\eta_j-\eta_{\tilde{B}})!(z'-\eta_B)!}{(\eta_j-\eta_{B-\{j\}})!(z'-\eta_{\tilde{B}})!}}~\alpha(z')\psi^\dagger_{z'}d^2z'|G^J\rangle\nonumber\\
&=&\frac{c_0}{N_J}\int_{z'\in S}\sum_{B'} \lambda^J_{B'}(-1)^{(j\in J)+\lfloor(j-1)/2\rfloor}\sqrt{\frac{(\eta_j-\eta_{\tilde{j}})(\eta_j-\eta_{\tilde{B'}})!(\eta_{\tilde{j}}-\eta_{\tilde{B'}})!(z'-\eta_{B'})!}{(z'-\eta_j)(z'-\eta_{\tilde{j}})(z'-\eta_{\tilde{B'}})!}}\alpha(z')\psi^\dagger_{z'}d^2z'|G^J\rangle,\nonumber
\end{eqnarray}
\end{widetext}
where $c_0=v_{0,j}(r_0)\sqrt{r_0}/\alpha(r_0) $ is an irrelevant constant, from the second to the third line we substituted $B=B'\cup \{j\}$ and the summation on the third line is over all $B'\subset S-\{j,\tilde{j}\}$ such that $B'$ contains exactly one of $\{i,\tilde{i}\}$ for $\forall i\in S,~i\neq j$, and we used the identity
\begin{equation}
\lambda^J_{B'\cup {j}}=\lambda^J_{B'}\sqrt{(\eta_j-\eta_{B'})!(\eta_{\tilde{j}}-\eta_{\tilde{B'}})!}~(-1)^{(j\in J)+\lfloor(j-1)/2\rfloor}
\end{equation}
which follows from the definition of $\lambda^J_B$ in Eq.~\eqref{eq:lambda^J_B}, the notation $(j\in J)$ is defined as $(P)=1$ if $P$ is true and $(P)=0$ if $P$ is false for an arbitrary statement $P$, and $\tilde{B'}=\{\tilde{j}~|~j\in B'\}$. From Eq.~\eqref{eq:gamma_jG^J} we obtain
\begin{equation}
[\gamma_{0,2j-1}+i(-1)^{(2j-1 \in J)}\gamma_{0,2j}]|G^J\rangle=0,
\end{equation}
which proves our claim that $|G^J\rangle$ is annihilated by $\chi_{j}$ if $(2j-1)\notin J$ and annihilated by $\chi_j^\dagger$ if $(2j-1)\in J$. Therefore,
$|G^J\rangle$ forms the occupation number basis
\begin{equation}
|G^J\rangle=C^J|n^J_1 n^J_2\ldots n^J_M\rangle,
\end{equation}
where $C^J$ is a normalization factor and $n^J_j=(2j-1\in J)$ is the eigenvalue of the operator $\hat{n}_j=\chi^\dagger_j\chi_j$.


\begin{thebibliography}{40}%
\makeatletter
\providecommand \@ifxundefined [1]{%
 \@ifx{#1\undefined}
}%
\providecommand \@ifnum [1]{%
 \ifnum #1\expandafter \@firstoftwo
 \else \expandafter \@secondoftwo
 \fi
}%
\providecommand \@ifx [1]{%
 \ifx #1\expandafter \@firstoftwo
 \else \expandafter \@secondoftwo
 \fi
}%
\providecommand \natexlab [1]{#1}%
\providecommand \enquote  [1]{``#1''}%
\providecommand \bibnamefont  [1]{#1}%
\providecommand \bibfnamefont [1]{#1}%
\providecommand \citenamefont [1]{#1}%
\providecommand \href@noop [0]{\@secondoftwo}%
\providecommand \href [0]{\begingroup \@sanitize@url \@href}%
\providecommand \@href[1]{\@@startlink{#1}\@@href}%
\providecommand \@@href[1]{\endgroup#1\@@endlink}%
\providecommand \@sanitize@url [0]{\catcode `\\12\catcode `\$12\catcode
  `\&12\catcode `\#12\catcode `\^12\catcode `\_12\catcode `\%12\relax}%
\providecommand \@@startlink[1]{}%
\providecommand \@@endlink[0]{}%
\providecommand \url  [0]{\begingroup\@sanitize@url \@url }%
\providecommand \@url [1]{\endgroup\@href {#1}{\urlprefix }}%
\providecommand \urlprefix  [0]{URL }%
\providecommand \Eprint [0]{\href }%
\providecommand \doibase [0]{http://dx.doi.org/}%
\providecommand \selectlanguage [0]{\@gobble}%
\providecommand \bibinfo  [0]{\@secondoftwo}%
\providecommand \bibfield  [0]{\@secondoftwo}%
\providecommand \translation [1]{[#1]}%
\providecommand \BibitemOpen [0]{}%
\providecommand \bibitemStop [0]{}%
\providecommand \bibitemNoStop [0]{.\EOS\space}%
\providecommand \EOS [0]{\spacefactor3000\relax}%
\providecommand \BibitemShut  [1]{\csname bibitem#1\endcsname}%
\let\auto@bib@innerbib\@empty
\bibitem [{\citenamefont {Qi}\ and\ \citenamefont {Zhang}(2011)}]{TPSC-RMP}%
  \BibitemOpen
  \bibfield  {author} {\bibinfo {author} {\bibfnamefont {X.-L.}\ \bibnamefont
  {Qi}}\ and\ \bibinfo {author} {\bibfnamefont {S.-C.}\ \bibnamefont {Zhang}},\
  }\href@noop {} {\bibfield  {journal} {\bibinfo  {journal} {Rev. Mod. Phys.}\
  }\textbf {\bibinfo {volume} {83}},\ \bibinfo {pages} {1057} (\bibinfo {year}
  {2011})}\BibitemShut {NoStop}%
\bibitem [{\citenamefont {Alicea}(2012)}]{Maj_CM1}%
  \BibitemOpen
  \bibfield  {author} {\bibinfo {author} {\bibfnamefont {J.}~\bibnamefont
  {Alicea}},\ }\href@noop {} {\bibfield  {journal} {\bibinfo  {journal} {Rep.
  Prog. Phys.}\ }\textbf {\bibinfo {volume} {75}},\ \bibinfo {pages} {076501}
  (\bibinfo {year} {2012})}\BibitemShut {NoStop}%
\bibitem [{\citenamefont {Beenakker}(2013)}]{Maj_CM2}%
  \BibitemOpen
  \bibfield  {author} {\bibinfo {author} {\bibfnamefont {C.~W.~J.}\
  \bibnamefont {Beenakker}},\ }\href@noop {} {\bibfield  {journal} {\bibinfo
  {journal} {Annu. Rev. Condens. Matter Phys.}\ }\textbf {\bibinfo {volume}
  {4}},\ \bibinfo {pages} {113} (\bibinfo {year} {2013})}\BibitemShut {NoStop}%
\bibitem [{\citenamefont {Zhang}\ \emph {et~al.}(2008)\citenamefont {Zhang},
  \citenamefont {Tewari}, \citenamefont {Lutchyn},\ and\ \citenamefont
  {Sarma}}]{Maj_coldatom1}%
  \BibitemOpen
  \bibfield  {author} {\bibinfo {author} {\bibfnamefont {C.}~\bibnamefont
  {Zhang}}, \bibinfo {author} {\bibfnamefont {S.}~\bibnamefont {Tewari}},
  \bibinfo {author} {\bibfnamefont {R.~M.}\ \bibnamefont {Lutchyn}}, \ and\
  \bibinfo {author} {\bibfnamefont {S.~D.}\ \bibnamefont {Sarma}},\ }\href@noop
  {} {\bibfield  {journal} {\bibinfo  {journal} {Phys. Rev. Lett.}\ }\textbf
  {\bibinfo {volume} {101}},\ \bibinfo {pages} {160401} (\bibinfo {year}
  {2008})}\BibitemShut {NoStop}%
\bibitem [{\citenamefont {Tewari}\ \emph {et~al.}(2007)\citenamefont {Tewari},
  \citenamefont {Sarma}, \citenamefont {Nayak}, \citenamefont {Zhang},\ and\
  \citenamefont {Zoller}}]{Maj_coldatom_QC}%
  \BibitemOpen
  \bibfield  {author} {\bibinfo {author} {\bibfnamefont {S.}~\bibnamefont
  {Tewari}}, \bibinfo {author} {\bibfnamefont {S.~D.}\ \bibnamefont {Sarma}},
  \bibinfo {author} {\bibfnamefont {C.}~\bibnamefont {Nayak}}, \bibinfo
  {author} {\bibfnamefont {C.}~\bibnamefont {Zhang}}, \ and\ \bibinfo {author}
  {\bibfnamefont {P.}~\bibnamefont {Zoller}},\ }\href@noop {} {\bibfield
  {journal} {\bibinfo  {journal} {Phys. Rev. Lett.}\ }\textbf {\bibinfo
  {volume} {98}},\ \bibinfo {pages} {010506} (\bibinfo {year}
  {2007})}\BibitemShut {NoStop}%
\bibitem [{\citenamefont {Moore}\ and\ \citenamefont {Read}(1991)}]{MR1991}%
  \BibitemOpen
  \bibfield  {author} {\bibinfo {author} {\bibfnamefont {G.}~\bibnamefont
  {Moore}}\ and\ \bibinfo {author} {\bibfnamefont {N.}~\bibnamefont {Read}},\
  }\href@noop {} {\bibfield  {journal} {\bibinfo  {journal} {Nucl. Phys. B}\
  }\textbf {\bibinfo {volume} {360}},\ \bibinfo {pages} {362} (\bibinfo {year}
  {1991})}\BibitemShut {NoStop}%
\bibitem [{\citenamefont {Nayak}\ and\ \citenamefont
  {Wilczek}(1996)}]{Nayak1996}%
  \BibitemOpen
  \bibfield  {author} {\bibinfo {author} {\bibfnamefont {C.}~\bibnamefont
  {Nayak}}\ and\ \bibinfo {author} {\bibfnamefont {F.}~\bibnamefont
  {Wilczek}},\ }\href@noop {} {\bibfield  {journal} {\bibinfo  {journal} {Nucl.
  Phys. B}\ }\textbf {\bibinfo {volume} {479}},\ \bibinfo {pages} {529}
  (\bibinfo {year} {1996})}\BibitemShut {NoStop}%
\bibitem [{\citenamefont {Ivanov}(2001)}]{Ivanov}%
  \BibitemOpen
  \bibfield  {author} {\bibinfo {author} {\bibfnamefont {D.~A.}\ \bibnamefont
  {Ivanov}},\ }\href@noop {} {\bibfield  {journal} {\bibinfo  {journal} {Phys.
  Rev. Lett.}\ }\textbf {\bibinfo {volume} {86}},\ \bibinfo {pages} {268}
  (\bibinfo {year} {2001})}\BibitemShut {NoStop}%
\bibitem [{\citenamefont {Moore}\ and\ \citenamefont
  {Seiberg}(1988)}]{Moore1988}%
  \BibitemOpen
  \bibfield  {author} {\bibinfo {author} {\bibfnamefont {G.}~\bibnamefont
  {Moore}}\ and\ \bibinfo {author} {\bibfnamefont {N.}~\bibnamefont
  {Seiberg}},\ }\href@noop {} {\bibfield  {journal} {\bibinfo  {journal} {Phys.
  Lett. B}\ }\textbf {\bibinfo {volume} {212}},\ \bibinfo {pages} {451}
  (\bibinfo {year} {1988})}\BibitemShut {NoStop}%
\bibitem [{\citenamefont {Witten}(1989)}]{Witten1989}%
  \BibitemOpen
  \bibfield  {author} {\bibinfo {author} {\bibfnamefont {E.}~\bibnamefont
  {Witten}},\ }\href@noop {} {\bibfield  {journal} {\bibinfo  {journal}
  {Commun. Math. Phys.}\ }\textbf {\bibinfo {volume} {121}},\ \bibinfo {pages}
  {351} (\bibinfo {year} {1989})}\BibitemShut {NoStop}%
\bibitem [{\citenamefont {Kitaev}(2003)}]{kitaevTC}%
  \BibitemOpen
  \bibfield  {author} {\bibinfo {author} {\bibfnamefont {A.~Y.}\ \bibnamefont
  {Kitaev}},\ }\href@noop {} {\bibfield  {journal} {\bibinfo  {journal} {Ann.
  Phys.}\ }\textbf {\bibinfo {volume} {303}},\ \bibinfo {pages} {2} (\bibinfo
  {year} {2003})}\BibitemShut {NoStop}%
\bibitem [{\citenamefont {Nayak}\ \emph {et~al.}(2008)\citenamefont {Nayak},
  \citenamefont {Simon}, \citenamefont {Stern}, \citenamefont {Freedman},\ and\
  \citenamefont {Das~Sarma}}]{TPQC-RMP}%
  \BibitemOpen
  \bibfield  {author} {\bibinfo {author} {\bibfnamefont {C.}~\bibnamefont
  {Nayak}}, \bibinfo {author} {\bibfnamefont {S.~H.}\ \bibnamefont {Simon}},
  \bibinfo {author} {\bibfnamefont {A.}~\bibnamefont {Stern}}, \bibinfo
  {author} {\bibfnamefont {M.}~\bibnamefont {Freedman}}, \ and\ \bibinfo
  {author} {\bibfnamefont {S.}~\bibnamefont {Das~Sarma}},\ }\href@noop {}
  {\bibfield  {journal} {\bibinfo  {journal} {Rev. Mod. Phys.}\ }\textbf
  {\bibinfo {volume} {80}},\ \bibinfo {pages} {1083} (\bibinfo {year}
  {2008})}\BibitemShut {NoStop}%
\bibitem [{\citenamefont {Mourik}\ \emph {et~al.}(2012)\citenamefont {Mourik},
  \citenamefont {Zuo}, \citenamefont {Frolov}, \citenamefont {Plissard},
  \citenamefont {Bakkers},\ and\ \citenamefont {Kouwenhoven}}]{Exp0}%
  \BibitemOpen
  \bibfield  {author} {\bibinfo {author} {\bibfnamefont {V.}~\bibnamefont
  {Mourik}}, \bibinfo {author} {\bibfnamefont {K.}~\bibnamefont {Zuo}},
  \bibinfo {author} {\bibfnamefont {S.~M.}\ \bibnamefont {Frolov}}, \bibinfo
  {author} {\bibfnamefont {S.}~\bibnamefont {Plissard}}, \bibinfo {author}
  {\bibfnamefont {E.}~\bibnamefont {Bakkers}}, \ and\ \bibinfo {author}
  {\bibfnamefont {L.~P.}\ \bibnamefont {Kouwenhoven}},\ }\href@noop {}
  {\bibfield  {journal} {\bibinfo  {journal} {Science}\ }\textbf {\bibinfo
  {volume} {336}},\ \bibinfo {pages} {1003} (\bibinfo {year}
  {2012})}\BibitemShut {NoStop}%
\bibitem [{\citenamefont {Deng}\ \emph {et~al.}(2012)\citenamefont {Deng},
  \citenamefont {Yu}, \citenamefont {Huang}, \citenamefont {Larsson},
  \citenamefont {Caroff},\ and\ \citenamefont {Xu}}]{Exp00}%
  \BibitemOpen
  \bibfield  {author} {\bibinfo {author} {\bibfnamefont {M.~T.}\ \bibnamefont
  {Deng}}, \bibinfo {author} {\bibfnamefont {C.~L.}\ \bibnamefont {Yu}},
  \bibinfo {author} {\bibfnamefont {G.~Y.}\ \bibnamefont {Huang}}, \bibinfo
  {author} {\bibfnamefont {M.}~\bibnamefont {Larsson}}, \bibinfo {author}
  {\bibfnamefont {P.}~\bibnamefont {Caroff}}, \ and\ \bibinfo {author}
  {\bibfnamefont {H.~Q.}\ \bibnamefont {Xu}},\ }\href@noop {} {\bibfield
  {journal} {\bibinfo  {journal} {Nano lett.}\ }\textbf {\bibinfo {volume}
  {12}},\ \bibinfo {pages} {6414} (\bibinfo {year} {2012})}\BibitemShut
  {NoStop}%
\bibitem [{\citenamefont {Nadj-Perge}\ \emph {et~al.}(2014)\citenamefont
  {Nadj-Perge}, \citenamefont {Drozdov}, \citenamefont {Li}, \citenamefont
  {Chen}, \citenamefont {Jeon}, \citenamefont {Seo}, \citenamefont {MacDonald},
  \citenamefont {Bernevig},\ and\ \citenamefont {Yazdani}}]{Exp1}%
  \BibitemOpen
  \bibfield  {author} {\bibinfo {author} {\bibfnamefont {S.}~\bibnamefont
  {Nadj-Perge}}, \bibinfo {author} {\bibfnamefont {I.~K.}\ \bibnamefont
  {Drozdov}}, \bibinfo {author} {\bibfnamefont {J.}~\bibnamefont {Li}},
  \bibinfo {author} {\bibfnamefont {H.}~\bibnamefont {Chen}}, \bibinfo {author}
  {\bibfnamefont {S.}~\bibnamefont {Jeon}}, \bibinfo {author} {\bibfnamefont
  {J.}~\bibnamefont {Seo}}, \bibinfo {author} {\bibfnamefont {A.~H.}\
  \bibnamefont {MacDonald}}, \bibinfo {author} {\bibfnamefont {B.~A.}\
  \bibnamefont {Bernevig}}, \ and\ \bibinfo {author} {\bibfnamefont
  {A.}~\bibnamefont {Yazdani}},\ }\href@noop {} {\bibfield  {journal} {\bibinfo
   {journal} {Science}\ }\textbf {\bibinfo {volume} {346}},\ \bibinfo {pages}
  {602} (\bibinfo {year} {2014})}\BibitemShut {NoStop}%
\bibitem [{\citenamefont {Xu}\ \emph {et~al.}(2015)\citenamefont {Xu},
  \citenamefont {Wang}, \citenamefont {Liu}, \citenamefont {Ge}, \citenamefont
  {Yang}, \citenamefont {Liu}, \citenamefont {Xu}, \citenamefont {Guan},
  \citenamefont {Gao}, \citenamefont {Qian} \emph {et~al.}}]{Exp2}%
  \BibitemOpen
  \bibfield  {author} {\bibinfo {author} {\bibfnamefont {J.-P.}\ \bibnamefont
  {Xu}}, \bibinfo {author} {\bibfnamefont {M.-X.}\ \bibnamefont {Wang}},
  \bibinfo {author} {\bibfnamefont {Z.~L.}\ \bibnamefont {Liu}}, \bibinfo
  {author} {\bibfnamefont {J.-F.}\ \bibnamefont {Ge}}, \bibinfo {author}
  {\bibfnamefont {X.}~\bibnamefont {Yang}}, \bibinfo {author} {\bibfnamefont
  {C.}~\bibnamefont {Liu}}, \bibinfo {author} {\bibfnamefont {Z.~A.}\
  \bibnamefont {Xu}}, \bibinfo {author} {\bibfnamefont {D.}~\bibnamefont
  {Guan}}, \bibinfo {author} {\bibfnamefont {C.~L.}\ \bibnamefont {Gao}},
  \bibinfo {author} {\bibfnamefont {D.}~\bibnamefont {Qian}},  \emph {et~al.},\
  }\href@noop {} {\bibfield  {journal} {\bibinfo  {journal} {Phys. Rev. Lett.}\
  }\textbf {\bibinfo {volume} {114}},\ \bibinfo {pages} {017001} (\bibinfo
  {year} {2015})}\BibitemShut {NoStop}%
\bibitem [{\citenamefont {Albrecht}\ \emph {et~al.}(2016)\citenamefont
  {Albrecht}, \citenamefont {Higginbotham}, \citenamefont {Madsen},
  \citenamefont {Kuemmeth}, \citenamefont {Jespersen}, \citenamefont
  {Nyg{\aa}rd}, \citenamefont {Krogstrup},\ and\ \citenamefont
  {Marcus}}]{Exp3}%
  \BibitemOpen
  \bibfield  {author} {\bibinfo {author} {\bibfnamefont {S.~M.}\ \bibnamefont
  {Albrecht}}, \bibinfo {author} {\bibfnamefont {A.}~\bibnamefont
  {Higginbotham}}, \bibinfo {author} {\bibfnamefont {M.}~\bibnamefont
  {Madsen}}, \bibinfo {author} {\bibfnamefont {F.}~\bibnamefont {Kuemmeth}},
  \bibinfo {author} {\bibfnamefont {T.~S.}\ \bibnamefont {Jespersen}}, \bibinfo
  {author} {\bibfnamefont {J.}~\bibnamefont {Nyg{\aa}rd}}, \bibinfo {author}
  {\bibfnamefont {P.}~\bibnamefont {Krogstrup}}, \ and\ \bibinfo {author}
  {\bibfnamefont {C.}~\bibnamefont {Marcus}},\ }\href@noop {} {\bibfield
  {journal} {\bibinfo  {journal} {Nature}\ }\textbf {\bibinfo {volume} {531}},\
  \bibinfo {pages} {206} (\bibinfo {year} {2016})}\BibitemShut {NoStop}%
\bibitem [{\citenamefont {Sun}\ \emph {et~al.}(2016)\citenamefont {Sun},
  \citenamefont {Zhang}, \citenamefont {Hu}, \citenamefont {Li}, \citenamefont
  {Wang}, \citenamefont {Ma}, \citenamefont {Xu}, \citenamefont {Gao},
  \citenamefont {Guan}, \citenamefont {Li} \emph {et~al.}}]{Exp4}%
  \BibitemOpen
  \bibfield  {author} {\bibinfo {author} {\bibfnamefont {H.-H.}\ \bibnamefont
  {Sun}}, \bibinfo {author} {\bibfnamefont {K.-W.}\ \bibnamefont {Zhang}},
  \bibinfo {author} {\bibfnamefont {L.-H.}\ \bibnamefont {Hu}}, \bibinfo
  {author} {\bibfnamefont {C.}~\bibnamefont {Li}}, \bibinfo {author}
  {\bibfnamefont {G.-Y.}\ \bibnamefont {Wang}}, \bibinfo {author}
  {\bibfnamefont {H.-Y.}\ \bibnamefont {Ma}}, \bibinfo {author} {\bibfnamefont
  {Z.-A.}\ \bibnamefont {Xu}}, \bibinfo {author} {\bibfnamefont {C.-L.}\
  \bibnamefont {Gao}}, \bibinfo {author} {\bibfnamefont {D.-D.}\ \bibnamefont
  {Guan}}, \bibinfo {author} {\bibfnamefont {Y.-Y.}\ \bibnamefont {Li}},  \emph
  {et~al.},\ }\href@noop {} {\bibfield  {journal} {\bibinfo  {journal} {Phys.
  Rev. Lett.}\ }\textbf {\bibinfo {volume} {116}},\ \bibinfo {pages} {257003}
  (\bibinfo {year} {2016})}\BibitemShut {NoStop}%
\bibitem [{\citenamefont {He}\ \emph {et~al.}(2017)\citenamefont {He},
  \citenamefont {Pan}, \citenamefont {Stern}, \citenamefont {Burks},
  \citenamefont {Che}, \citenamefont {Yin}, \citenamefont {Wang}, \citenamefont
  {Lian}, \citenamefont {Zhou}, \citenamefont {Choi} \emph
  {et~al.}}]{ExpZhang}%
  \BibitemOpen
  \bibfield  {author} {\bibinfo {author} {\bibfnamefont {Q.~L.}\ \bibnamefont
  {He}}, \bibinfo {author} {\bibfnamefont {L.}~\bibnamefont {Pan}}, \bibinfo
  {author} {\bibfnamefont {A.~L.}\ \bibnamefont {Stern}}, \bibinfo {author}
  {\bibfnamefont {E.~C.}\ \bibnamefont {Burks}}, \bibinfo {author}
  {\bibfnamefont {X.}~\bibnamefont {Che}}, \bibinfo {author} {\bibfnamefont
  {G.}~\bibnamefont {Yin}}, \bibinfo {author} {\bibfnamefont {J.}~\bibnamefont
  {Wang}}, \bibinfo {author} {\bibfnamefont {B.}~\bibnamefont {Lian}}, \bibinfo
  {author} {\bibfnamefont {Q.}~\bibnamefont {Zhou}}, \bibinfo {author}
  {\bibfnamefont {E.~S.}\ \bibnamefont {Choi}},  \emph {et~al.},\ }\href@noop
  {} {\bibfield  {journal} {\bibinfo  {journal} {Science}\ }\textbf {\bibinfo
  {volume} {357}},\ \bibinfo {pages} {294} (\bibinfo {year}
  {2017})}\BibitemShut {NoStop}%
\bibitem [{\citenamefont {Kitaev}(2001)}]{Kitaev}%
  \BibitemOpen
  \bibfield  {author} {\bibinfo {author} {\bibfnamefont {A.~Y.}\ \bibnamefont
  {Kitaev}},\ }\href@noop {} {\bibfield  {journal} {\bibinfo  {journal}
  {Phys.-Usp.}\ }\textbf {\bibinfo {volume} {44}},\ \bibinfo {pages} {131}
  (\bibinfo {year} {2001})}\BibitemShut {NoStop}%
\bibitem [{\citenamefont {Greiter}\ \emph {et~al.}(2014)\citenamefont
  {Greiter}, \citenamefont {Schnells},\ and\ \citenamefont
  {Thomale}}]{AoP2014}%
  \BibitemOpen
  \bibfield  {author} {\bibinfo {author} {\bibfnamefont {M.}~\bibnamefont
  {Greiter}}, \bibinfo {author} {\bibfnamefont {V.}~\bibnamefont {Schnells}}, \
  and\ \bibinfo {author} {\bibfnamefont {R.}~\bibnamefont {Thomale}},\
  }\href@noop {} {\bibfield  {journal} {\bibinfo  {journal} {Ann. Phys.}\
  }\textbf {\bibinfo {volume} {351}},\ \bibinfo {pages} {1026} (\bibinfo {year}
  {2014})}\BibitemShut {NoStop}%
\bibitem [{\citenamefont {Alicea}\ \emph {et~al.}(2010)\citenamefont {Alicea},
  \citenamefont {Oreg}, \citenamefont {Refael}, \citenamefont {Von~Oppen},\
  and\ \citenamefont {Fisher}}]{Alicea2010}%
  \BibitemOpen
  \bibfield  {author} {\bibinfo {author} {\bibfnamefont {J.}~\bibnamefont
  {Alicea}}, \bibinfo {author} {\bibfnamefont {Y.}~\bibnamefont {Oreg}},
  \bibinfo {author} {\bibfnamefont {G.}~\bibnamefont {Refael}}, \bibinfo
  {author} {\bibfnamefont {F.}~\bibnamefont {Von~Oppen}}, \ and\ \bibinfo
  {author} {\bibfnamefont {M.}~\bibnamefont {Fisher}},\ }\href@noop {}
  {\bibfield  {journal} {\bibinfo  {journal} {arXiv preprint arXiv:1006.4395}\
  } (\bibinfo {year} {2010})}\BibitemShut {NoStop}%
\bibitem [{\citenamefont {Read}\ and\ \citenamefont {Green}(2000)}]{NRead}%
  \BibitemOpen
  \bibfield  {author} {\bibinfo {author} {\bibfnamefont {N.}~\bibnamefont
  {Read}}\ and\ \bibinfo {author} {\bibfnamefont {D.}~\bibnamefont {Green}},\
  }\href@noop {} {\bibfield  {journal} {\bibinfo  {journal} {Phys. Rev. B}\
  }\textbf {\bibinfo {volume} {61}},\ \bibinfo {pages} {10267} (\bibinfo {year}
  {2000})}\BibitemShut {NoStop}%
\bibitem [{\citenamefont {Wang}\ \emph {et~al.}(2017)\citenamefont {Wang},
  \citenamefont {Xu}, \citenamefont {Pu},\ and\ \citenamefont
  {Hazzard}}]{ZWang}%
  \BibitemOpen
  \bibfield  {author} {\bibinfo {author} {\bibfnamefont {Z.}~\bibnamefont
  {Wang}}, \bibinfo {author} {\bibfnamefont {Y.}~\bibnamefont {Xu}}, \bibinfo
  {author} {\bibfnamefont {H.}~\bibnamefont {Pu}}, \ and\ \bibinfo {author}
  {\bibfnamefont {K.~R.}\ \bibnamefont {Hazzard}},\ }\href@noop {} {\bibfield
  {journal} {\bibinfo  {journal} {Phys. Rev. B}\ }\textbf {\bibinfo {volume}
  {96}},\ \bibinfo {pages} {115110} (\bibinfo {year} {2017})}\BibitemShut
  {NoStop}%
\bibitem [{\citenamefont {Iemini}\ \emph {et~al.}(2015)\citenamefont {Iemini},
  \citenamefont {Mazza}, \citenamefont {Rossini}, \citenamefont {Fazio},\ and\
  \citenamefont {Diehl}}]{SDiehl}%
  \BibitemOpen
  \bibfield  {author} {\bibinfo {author} {\bibfnamefont {F.}~\bibnamefont
  {Iemini}}, \bibinfo {author} {\bibfnamefont {L.}~\bibnamefont {Mazza}},
  \bibinfo {author} {\bibfnamefont {D.}~\bibnamefont {Rossini}}, \bibinfo
  {author} {\bibfnamefont {R.}~\bibnamefont {Fazio}}, \ and\ \bibinfo {author}
  {\bibfnamefont {S.}~\bibnamefont {Diehl}},\ }\href@noop {} {\bibfield
  {journal} {\bibinfo  {journal} {Phys. Rev. Lett.}\ }\textbf {\bibinfo
  {volume} {115}},\ \bibinfo {pages} {156402} (\bibinfo {year}
  {2015})}\BibitemShut {NoStop}%
\bibitem [{\citenamefont {Lang}\ and\ \citenamefont
  {B{\"u}chler}(2015)}]{NLang}%
  \BibitemOpen
  \bibfield  {author} {\bibinfo {author} {\bibfnamefont {N.}~\bibnamefont
  {Lang}}\ and\ \bibinfo {author} {\bibfnamefont {H.~P.}\ \bibnamefont
  {B{\"u}chler}},\ }\href@noop {} {\bibfield  {journal} {\bibinfo  {journal}
  {Phys. Rev. B}\ }\textbf {\bibinfo {volume} {92}},\ \bibinfo {pages} {041118}
  (\bibinfo {year} {2015})}\BibitemShut {NoStop}%
\bibitem [{\citenamefont {Affleck}\ \emph {et~al.}(1988)\citenamefont
  {Affleck}, \citenamefont {Kennedy}, \citenamefont {Lieb},\ and\ \citenamefont
  {Tasaki}}]{AKLT}%
  \BibitemOpen
  \bibfield  {author} {\bibinfo {author} {\bibfnamefont {I.}~\bibnamefont
  {Affleck}}, \bibinfo {author} {\bibfnamefont {T.}~\bibnamefont {Kennedy}},
  \bibinfo {author} {\bibfnamefont {E.~H.}\ \bibnamefont {Lieb}}, \ and\
  \bibinfo {author} {\bibfnamefont {H.}~\bibnamefont {Tasaki}},\ }\href@noop {}
  {\bibfield  {journal} {\bibinfo  {journal} {Commun. Math. Phys.}\ }\textbf
  {\bibinfo {volume} {115}},\ \bibinfo {pages} {477} (\bibinfo {year}
  {1988})}\BibitemShut {NoStop}%
\bibitem [{\citenamefont {Rokhsar}\ and\ \citenamefont
  {Kivelson}(1988)}]{RKmodel}%
  \BibitemOpen
  \bibfield  {author} {\bibinfo {author} {\bibfnamefont {D.~S.}\ \bibnamefont
  {Rokhsar}}\ and\ \bibinfo {author} {\bibfnamefont {S.~A.}\ \bibnamefont
  {Kivelson}},\ }\href@noop {} {\bibfield  {journal} {\bibinfo  {journal}
  {Phys. Rev. Lett}\ }\textbf {\bibinfo {volume} {61}},\ \bibinfo {pages}
  {2376} (\bibinfo {year} {1988})}\BibitemShut {NoStop}%
\bibitem [{\citenamefont {Rombouts}\ \emph {et~al.}(2010)\citenamefont
  {Rombouts}, \citenamefont {Dukelsky},\ and\ \citenamefont
  {Ortiz}}]{Dukelsky2010}%
  \BibitemOpen
  \bibfield  {author} {\bibinfo {author} {\bibfnamefont {S.~M.~A.}\
  \bibnamefont {Rombouts}}, \bibinfo {author} {\bibfnamefont {J.}~\bibnamefont
  {Dukelsky}}, \ and\ \bibinfo {author} {\bibfnamefont {G.}~\bibnamefont
  {Ortiz}},\ }\href@noop {} {\bibfield  {journal} {\bibinfo  {journal} {Phys.
  Rev. B}\ }\textbf {\bibinfo {volume} {82}},\ \bibinfo {pages} {224510}
  (\bibinfo {year} {2010})}\BibitemShut {NoStop}%
\bibitem [{\citenamefont {Dunning}\ \emph {et~al.}(2010)\citenamefont
  {Dunning}, \citenamefont {Ibanez}, \citenamefont {Links}, \citenamefont
  {Sierra},\ and\ \citenamefont {Zhao}}]{JSM2010}%
  \BibitemOpen
  \bibfield  {author} {\bibinfo {author} {\bibfnamefont {C.}~\bibnamefont
  {Dunning}}, \bibinfo {author} {\bibfnamefont {M.}~\bibnamefont {Ibanez}},
  \bibinfo {author} {\bibfnamefont {J.}~\bibnamefont {Links}}, \bibinfo
  {author} {\bibfnamefont {G.}~\bibnamefont {Sierra}}, \ and\ \bibinfo {author}
  {\bibfnamefont {S.-Y.}\ \bibnamefont {Zhao}},\ }\href@noop {} {\bibfield
  {journal} {\bibinfo  {journal} {J. Stat. Mech.}\ ,\ \bibinfo {pages}
  {P08025}} (\bibinfo {year} {2010})}\BibitemShut {NoStop}%
\bibitem [{\citenamefont {Greiter}\ \emph {et~al.}(1991)\citenamefont
  {Greiter}, \citenamefont {Wen},\ and\ \citenamefont {Wilczek}}]{Greiter1991}%
  \BibitemOpen
  \bibfield  {author} {\bibinfo {author} {\bibfnamefont {M.}~\bibnamefont
  {Greiter}}, \bibinfo {author} {\bibfnamefont {X.-G.}\ \bibnamefont {Wen}}, \
  and\ \bibinfo {author} {\bibfnamefont {F.}~\bibnamefont {Wilczek}},\
  }\href@noop {} {\bibfield  {journal} {\bibinfo  {journal} {Phys. Rev. Lett.}\
  }\textbf {\bibinfo {volume} {66}},\ \bibinfo {pages} {3205} (\bibinfo {year}
  {1991})}\BibitemShut {NoStop}%
\bibitem [{\citenamefont {Greiter}\ \emph {et~al.}(1992)\citenamefont
  {Greiter}, \citenamefont {Wen},\ and\ \citenamefont {Wilczek}}]{Greiter1992}%
  \BibitemOpen
  \bibfield  {author} {\bibinfo {author} {\bibfnamefont {M.}~\bibnamefont
  {Greiter}}, \bibinfo {author} {\bibfnamefont {X.-G.}\ \bibnamefont {Wen}}, \
  and\ \bibinfo {author} {\bibfnamefont {F.}~\bibnamefont {Wilczek}},\
  }\href@noop {} {\bibfield  {journal} {\bibinfo  {journal} {Nucl. Phys. B}\
  }\textbf {\bibinfo {volume} {374}},\ \bibinfo {pages} {567} (\bibinfo {year}
  {1992})}\BibitemShut {NoStop}%
\bibitem [{\citenamefont {Read}\ and\ \citenamefont
  {Rezayi}(1996)}]{Rezayi1996}%
  \BibitemOpen
  \bibfield  {author} {\bibinfo {author} {\bibfnamefont {N.}~\bibnamefont
  {Read}}\ and\ \bibinfo {author} {\bibfnamefont {E.}~\bibnamefont {Rezayi}},\
  }\href@noop {} {\bibfield  {journal} {\bibinfo  {journal} {Phys. Rev. B}\
  }\textbf {\bibinfo {volume} {54}},\ \bibinfo {pages} {16864} (\bibinfo {year}
  {1996})}\BibitemShut {NoStop}%
\bibitem [{\citenamefont {Katsura}\ \emph {et~al.}(2015)\citenamefont
  {Katsura}, \citenamefont {Schuricht},\ and\ \citenamefont
  {Takahashi}}]{InteractingKitaev}%
  \BibitemOpen
  \bibfield  {author} {\bibinfo {author} {\bibfnamefont {H.}~\bibnamefont
  {Katsura}}, \bibinfo {author} {\bibfnamefont {D.}~\bibnamefont {Schuricht}},
  \ and\ \bibinfo {author} {\bibfnamefont {M.}~\bibnamefont {Takahashi}},\
  }\href@noop {} {\bibfield  {journal} {\bibinfo  {journal} {Phys. Rev. B}\
  }\textbf {\bibinfo {volume} {92}},\ \bibinfo {pages} {115137} (\bibinfo
  {year} {2015})}\BibitemShut {NoStop}%
\bibitem [{\citenamefont {Bonderson}\ \emph {et~al.}(2011)\citenamefont
  {Bonderson}, \citenamefont {Gurarie},\ and\ \citenamefont
  {Nayak}}]{Bonderson2011}%
  \BibitemOpen
  \bibfield  {author} {\bibinfo {author} {\bibfnamefont {P.}~\bibnamefont
  {Bonderson}}, \bibinfo {author} {\bibfnamefont {V.}~\bibnamefont {Gurarie}},
  \ and\ \bibinfo {author} {\bibfnamefont {C.}~\bibnamefont {Nayak}},\
  }\href@noop {} {\bibfield  {journal} {\bibinfo  {journal} {Phys. Rev. B}\
  }\textbf {\bibinfo {volume} {83}},\ \bibinfo {pages} {075303} (\bibinfo
  {year} {2011})}\BibitemShut {NoStop}%
\bibitem [{\citenamefont {Fidkowski}\ \emph {et~al.}(2011)\citenamefont
  {Fidkowski}, \citenamefont {Lutchyn}, \citenamefont {Nayak},\ and\
  \citenamefont {Fisher}}]{Bos1}%
  \BibitemOpen
  \bibfield  {author} {\bibinfo {author} {\bibfnamefont {L.}~\bibnamefont
  {Fidkowski}}, \bibinfo {author} {\bibfnamefont {R.~M.}\ \bibnamefont
  {Lutchyn}}, \bibinfo {author} {\bibfnamefont {C.}~\bibnamefont {Nayak}}, \
  and\ \bibinfo {author} {\bibfnamefont {M.~P.~A.}\ \bibnamefont {Fisher}},\
  }\href@noop {} {\bibfield  {journal} {\bibinfo  {journal} {Phys. Rev. B}\
  }\textbf {\bibinfo {volume} {84}},\ \bibinfo {pages} {195436} (\bibinfo
  {year} {2011})}\BibitemShut {NoStop}%
\bibitem [{\citenamefont {Sau}\ \emph {et~al.}(2011)\citenamefont {Sau},
  \citenamefont {Halperin}, \citenamefont {Flensberg},\ and\ \citenamefont
  {Das~Sarma}}]{Bos2}%
  \BibitemOpen
  \bibfield  {author} {\bibinfo {author} {\bibfnamefont {J.~D.}\ \bibnamefont
  {Sau}}, \bibinfo {author} {\bibfnamefont {B.~I.}\ \bibnamefont {Halperin}},
  \bibinfo {author} {\bibfnamefont {K.}~\bibnamefont {Flensberg}}, \ and\
  \bibinfo {author} {\bibfnamefont {S.}~\bibnamefont {Das~Sarma}},\ }\href@noop
  {} {\bibfield  {journal} {\bibinfo  {journal} {Phys. Rev. B}\ }\textbf
  {\bibinfo {volume} {84}},\ \bibinfo {pages} {144509} (\bibinfo {year}
  {2011})}\BibitemShut {NoStop}%
\bibitem [{\citenamefont {Cheng}\ and\ \citenamefont {Tu}(2011)}]{Bos3}%
  \BibitemOpen
  \bibfield  {author} {\bibinfo {author} {\bibfnamefont {M.}~\bibnamefont
  {Cheng}}\ and\ \bibinfo {author} {\bibfnamefont {H.-H.}\ \bibnamefont {Tu}},\
  }\href@noop {} {\bibfield  {journal} {\bibinfo  {journal} {Phys. Rev. B}\
  }\textbf {\bibinfo {volume} {84}},\ \bibinfo {pages} {094503} (\bibinfo
  {year} {2011})}\BibitemShut {NoStop}%
\bibitem [{\citenamefont {Kraus}\ \emph {et~al.}(2013)\citenamefont {Kraus},
  \citenamefont {Dalmonte}, \citenamefont {Baranov}, \citenamefont
  {L{\"a}uchli},\ and\ \citenamefont {Zoller}}]{PZoller}%
  \BibitemOpen
  \bibfield  {author} {\bibinfo {author} {\bibfnamefont {C.~V.}\ \bibnamefont
  {Kraus}}, \bibinfo {author} {\bibfnamefont {M.}~\bibnamefont {Dalmonte}},
  \bibinfo {author} {\bibfnamefont {M.~A.}\ \bibnamefont {Baranov}}, \bibinfo
  {author} {\bibfnamefont {A.~M.}\ \bibnamefont {L{\"a}uchli}}, \ and\ \bibinfo
  {author} {\bibfnamefont {P.}~\bibnamefont {Zoller}},\ }\href@noop {}
  {\bibfield  {journal} {\bibinfo  {journal} {Phys. Rev. Lett.}\ }\textbf
  {\bibinfo {volume} {111}},\ \bibinfo {pages} {173004} (\bibinfo {year}
  {2013})}\BibitemShut {NoStop}%
\bibitem [{\citenamefont {Ortiz}\ \emph {et~al.}(2014)\citenamefont {Ortiz},
  \citenamefont {Dukelsky}, \citenamefont {Cobanera}, \citenamefont {Esebbag},\
  and\ \citenamefont {Beenakker}}]{RGK}%
  \BibitemOpen
  \bibfield  {author} {\bibinfo {author} {\bibfnamefont {G.}~\bibnamefont
  {Ortiz}}, \bibinfo {author} {\bibfnamefont {J.}~\bibnamefont {Dukelsky}},
  \bibinfo {author} {\bibfnamefont {E.}~\bibnamefont {Cobanera}}, \bibinfo
  {author} {\bibfnamefont {C.}~\bibnamefont {Esebbag}}, \ and\ \bibinfo
  {author} {\bibfnamefont {C.}~\bibnamefont {Beenakker}},\ }\href@noop {}
  {\bibfield  {journal} {\bibinfo  {journal} {Phys. Rev. Lett.}\ }\textbf
  {\bibinfo {volume} {113}},\ \bibinfo {pages} {267002} (\bibinfo {year}
  {2014})}\BibitemShut {NoStop}%
\end{thebibliography}
\end{document}